\documentclass[11pt,a4paper]{article}
\usepackage{cite}
\usepackage{graphicx}
\usepackage{amssymb}
\usepackage{amsmath}
\usepackage{amsfonts}
\usepackage{dsfont}
\usepackage{mathtools}
\usepackage{array}
\usepackage{rotating}
\usepackage{bbold,amsfonts}

\usepackage[utf8]{inputenc}
\usepackage{bm}
\usepackage{xcolor}
\usepackage{float}
\usepackage{braket}
\usepackage{makecell}

\usepackage[height=8.8in,width=6.45in]{geometry}
\usepackage[font=small,labelfont=bf]{caption}
\usepackage[hidelinks]{hyperref}
\usepackage{booktabs,float,slashed}
\usepackage{standalone}
\usepackage{caption}
\usepackage{subcaption}
\numberwithin{equation}{section}

\newcommand{\beq}{\begin{equation}}
\newcommand{\eeq}{\end{equation}}

\newcommand{\overbar}[1]{\mkern 1.5mu\overline{\mkern-1.5mu#1\mkern-1.5mu}\mkern 1.5mu}

\makeatletter
\newcommand*{\letterdef@}{}
\newcommand*{\letterdef}[3]{%
	\def\letterdef@##1{\expandafter\newcommand\csname #1\endcsname{#2{##1}}}%
	\@tfor\@tempa :=#3\do{\expandafter\letterdef@\expandafter{\@tempa}}}
\makeatother
\letterdef{c#1} {\mathcal}{ABCDEFGHIJKLMNOPQRSTUVWXYZ} 
\letterdef{rm#1}{\mathrm} {dDeimM} 

\newdimen\tableauside\tableauside=1.0ex
\newdimen\tableaurule\tableaurule=0.4pt
\newdimen\tableaustep
\def\phantomhrule#1{\hbox{\vbox to0pt{\hrule height\tableaurule
			width#1\vss}}}
\def\phantomvrule#1{\vbox{\hbox to0pt{\vrule width\tableaurule
			height#1\hss}}}
\def\sqr{\vbox{%
		\phantomhrule\tableaustep
		\hbox{\phantomvrule\tableaustep\kern\tableaustep\phantomvrule\tableaustep}%
		\hbox{\vbox{\phantomhrule\tableauside}\kern-\tableaurule}}}
\def\squares#1{\hbox{\count0=#1\noindent\loop\sqr
		\advance\count0 by-1 \ifnum\count0>0\repeat}}
\def\tableau#1{\vcenter{\offinterlineskip
		\tableaustep=\tableauside\advance\tableaustep by-\tableaurule
		\kern\normallineskip\hbox
		{\kern\normallineskip\vbox
			{\gettableau#1 0 }%
			\kern\normallineskip\kern\tableaurule}%
		\kern\normallineskip\kern\tableaurule}}
\def\gettableau#1 {\ifnum#1=0\let\next=\null\else
	\squares{#1}\let\next=\gettableau\fi\next}
\allowdisplaybreaks

\begin{document}
\begin{titlepage}
\vspace*{10mm}
\begin{center}
{\LARGE \bf 
Strong coupling expansions in \\[1mm]
$\mathcal{N}=2$ quiver gauge theories
}

\vspace*{15mm}

{\Large M. Bill\`o${}^{\,a,c}$, M. Frau${}^{\,a,c}$, A. Lerda${}^{\,b,c}$, A. Pini${}^{\,a,c}$, P. Vallarino${}^{\,a,c}$}

\vspace*{8mm}
	
${}^a$ Universit\`a di Torino, Dipartimento di Fisica,\\
			Via P. Giuria 1, I-10125 Torino, Italy
			\vskip 0.3cm

${}^b$  Universit\`a del Piemonte Orientale,\\
			Dipartimento di Scienze e Innovazione Tecnologica\\
			Viale T. Michel 11, I-15121 Alessandria, Italy
			\vskip 0.3cm
			
${}^c$   I.N.F.N. - sezione di Torino,\\
			Via P. Giuria 1, I-10125 Torino, Italy

\vskip 0.8cm
	{\small
		E-mail:
		\texttt{billo,frau,lerda,apini,vallarin@to.infn.it}
	}
\vspace*{0.8cm}
\end{center}

\begin{abstract}

We study the 3-point functions of gauge-invariant scalar operators in four dimensional $\mathcal{N}=2$ superconformal quiver theories using supersymmetric localization in the planar limit of a large number of colors. By exploiting a web of nontrivial relations, we show that the 3-point functions can be expressed in terms of the 2-point functions through exact Ward-like identities that are valid for all values of the coupling constant. In this way, using recent results about the 2-point functions, we are able to obtain the asymptotic strong-coupling expansion of the 3-point functions and of the corresponding structure constants in the planar limit. Our results extend to sub-leading orders what has been recently found at leading order, where a precise match with calculations within the AdS/CFT correspondence at the supergravity level is possible. Therefore, our findings can be interpreted also as a prediction for the sub-leading string corrections to these holographic calculations. 

\end{abstract}
\end{titlepage}
\setcounter{tocdepth}{2}
\tableofcontents
\vspace{1cm}

\section{Introduction}
\label{sec:intro}
The analysis of the strong-coupling regime in an interacting theory is notoriously a very difficult problem but, when there is a high amount of symmetry, significant progress can be made. 
This is the case, for example, of the maximally supersymmetric SU($N$) Yang-Mills theory in four dimensions ($\mathcal{N}=4$ SYM) where many strong-coupling results can be obtained, especially in the planar limit of a large number of colors. The primary tool used to achieve this is the holographic AdS/CFT correspondence \cite{Maldacena:1997re} that maps strong-coupling calculations in 
$\mathcal{N}=4$ SYM into perturbative gravitational calculations in an AdS$_5\times S^5$ space-time \cite{Gubser:1998bc,Witten:1998qj}. 

Another tool that has been widely used at strong coupling is supersymmetric localization \cite{Pestun:2007rz}\,%
\footnote{For a review see, for instance, \cite{Pestun:2016jze}.}, which, by reducing path integrals to finite dimensional integrals, often yields expressions that are exact in the coupling constant. As an example of this, we mention the integrated correlators of four superconformal primaries of $\mathcal{N}=4$ SYM \cite{Binder:2019jwn,Chester:2020dja} whose exact properties have been explored in \cite{Dorigoni:2021guq}. Interestingly, the strong-coupling expansion of these integrated correlators, combined with suggestions from the structure of the four-graviton amplitude in string theory, can be used to test the AdS/CFT correspondence beyond the supergravity approximation \cite{Binder:2019jwn}. 

Other useful methods for the study of strongly coupled $\mathcal{N}=4$ SYM are provided by the integrability of the theory in the planar limit. For example, the correlation functions of four very heavy single-trace scalar operators can be factorized in terms of ``octagon'' form-factors \cite{Coronado:2018ypq,Coronado:2018cxj,Belitsky:2019fan,Belitsky:2020qrm,Belitsky:2020qir} whose properties are suggested by the integrability of the two-dimensional world-sheet in the AdS/CFT correspondence. Exploiting their representation as a Fredholm determinant of a Bessel operator, the octagons can be systematically studied at strong coupling and an explicit expansion in inverse powers of the coupling constant can be worked out \cite{Belitsky:2019fan,Belitsky:2020qrm,Belitsky:2020qir}. According to the holographic dictionary, the terms in this expansion originate from the string corrections to the supergravity effective action. However, a derivation of these corrections by a holographic calculation beyond the supergravity limit seems to be out of reach and thus the strong-coupling expansion derived from the Bessel operator is practically the only way to obtain information about the octagon form-factors and the corresponding correlators when the coupling constant is large.

The examples just mentioned refer to $\mathcal{N}=4$ SYM but, recently, there have been important developments along these lines also in theories with reduced supersymmetry. In particular, much attention has been paid to the four-dimensional $\mathcal{N}=2$ theories that arise as a
$\mathbb{Z}_M$ orbifold of $\mathcal{N}=4$ SYM. These theories have a quiver-like structure with $M$ nodes: their gauge group is the product of $M$ SU($N$) factors and there are bi-fundamental matter hypermultiplets. They possess a holographic dual given by Type II B string theory on a space-time of the type AdS$_5\times S^5/\mathbb{Z}_M$ \cite{Kachru:1998ys,Gukov:1998kk}. For these reasons they represent one of the simplest contexts in which to investigate the strong-coupling regime and explore the holographic correspondence when supersymmetry is not maximal. Applying supersymmetric localization to these
$\mathcal{N}=2$ quiver theories it has been possible to study several BPS-protected observables even at strong coupling. Among such observables we mention the free energy and the vacuum-expectation-value of the circular Wilson loop \cite{Zarembo:2020tpf,Fiol:2020ojn,Ouyang:2020hwd,Beccaria:2021ksw,Beccaria:2021vuc,Galvagno:2021bbj,Beccaria:2021ism} for which a systematic strong-coupling expansion has been recently worked out in \cite{Beccaria:2022ypy}. 

Also the 2- and 3-point functions of single-trace scalar operators of the quiver theories can be studied at strong coupling. In fact,
building on previous works that developed matrix-model techniques to study correlations functions in superconformal gauge theories \cite{Baggio:2014sna,Baggio:2015vxa,Gerchkovitz:2016gxx,Baggio:2016skg,Rodriguez-Gomez:2016ijh,Rodriguez-Gomez:2016cem,Pini:2017ouj,Billo:2017glv,Billo:2019fbi,Beccaria:2020azj,Beccaria:2020hgy,Galvagno:2020cgq,Beccaria:2021hvt,Fiol:2021icm,Billo:2021rdb,Billo:2022xas},
it has been shown in \cite{Billo:2022gmq,Billo:2022fnb} that the 2- and 3-point functions of single-trace scalar operators in the planar limit can be written in terms of the elements of a Bessel operator at all values of the coupling constant. Exploiting these results, it is possible to find in an analytic way the leading behavior at strong coupling of these correlators and show that the corresponding normalized structure constants precisely agree with the predictions of the AdS/CFT correspondence in a $\mathbb{Z}_M$ orbifold background \cite{Billo:2022gmq,Billo:2022fnb}. This agreement represents a highly non-trivial test of holography in $\mathcal{N}=2$ superconformal gauge theories at leading order.
To go further, some information on the sub-leading corrections is clearly necessary.

Actually, for the 2-point functions a systematic strong-coupling expansion has been recently obtained in
\cite{Beccaria:2022ypy} using analytic methods similar to those applied for the analysis of the octagons in $\mathcal{N}=4$ SYM. On the other hand, a high-precision numerical approach developed in \cite{Bobev:2022grf} has allowed to infer the first two sub-leading corrections to the 3-point correlators in the so-called $\mathbf{E}$ theory \cite{Billo:2019fbi,Beccaria:2021hvt,Billo:2022xas} which is obtained from the $M=2$ quiver theory by means of an orientifold projection. However,  
a systematic analytic expansion for the 3-point functions and the structure constants at strong coupling 
is still missing.

The purpose of this paper is to fill this gap. We do this by proving an exact relation that allows to express
the 3-point functions and the structure constants in terms of the 2-point functions for any value of the 't Hooft coupling. In this way, by exploiting the known behavior of the 2-point functions at strong coupling, we are able to deduce in an analytic way the strong-coupling expansion of the 3-point functions and of the structure constants, and to exhibit the first few sub-leading terms in a closed form. At the moment this is the only available analytic method to investigate the strong-coupling regime of the quiver theories beyond the leading order. Indeed, holographic calculations of 2- and 3-point correlators and of the structure constants beyond the supergravity approximation do not seem possible with the existent toolkit of the AdS/CFT correspondence, since they would require a detailed knowledge of the higher-derivative string corrections to the supergravity effective action in all twisted and untwisted sectors of the AdS$_5\times S^5/\mathbb{Z}_M$ orbifold.

In the following, with the aim of discussing our results in the simplest case, we will limit our analysis to the 2-node quiver ({\it{i.e.}} $M=2$) whose main features are recalled in Section~\ref{sec:quiver}. Then, in Section~\ref{sec:localization}
we briefly review the essential ingredients of the localization procedure that are used to compute the 2- and 3-point functions in the quiver theory.
In Section~\ref{sec:2point}, building on the results of our previous works \cite{Billo:2022gmq,Billo:2022fnb} and of \cite{Beccaria:2022ypy}, we derive the strong-coupling expansion of the 2-point correlators. We then proceed to study the 3-point functions by firstly considering in Section~\ref{sec:diagonal} a special family of 3-point functions whose strong-coupling expansion can be obtained from that of the 2-point functions to which they are related by a differential relation. Such a relation is instrumental for the calculations described in Section~\ref{sec:general} where in full generality we show that the 3-point functions and the structure constants can be written in terms of the 2-point functions. Finally in Section~\ref{sec:conclusions} we draw our conclusions, while in the appendices we collect some more technical details. In particular in Appendix~\ref{appendix:B} we describe an alternative way to obtain the strong-coupling expansion of the 3-point functions based on a recursive bootstrap-like procedure.

\section{The \texorpdfstring{$\mathbb{Z}_2$}{} quiver theory}
\label{sec:quiver}
\begin{figure}[ht]
    \centering
    \includegraphics[scale=0.75]{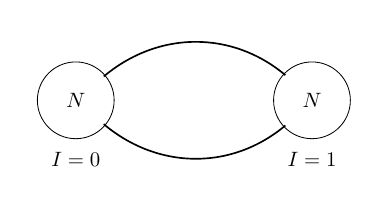}
    \caption{The $4d$ $\mathcal{N}=2$ quiver gauge theory.}
    \label{fig:quiver}
\end{figure}

We consider the 2-node $\mathcal{N}=2$ quiver theory obtained with a $\mathbb{Z}_2$ orbifold projection from $\mathcal{N}=4$ SYM in four dimensions. This theory is schematically represented by the diagram in 
Fig.~\ref{fig:quiver}.
Here each node, labeled by the index $I=0,1$, corresponds to a SU$(N)$ gauge group with its adjoint vector multiplet while the lines connecting the nodes represent $\mathcal{N}=2$ hypermultiplets in the bi-fundamental representation.
Given this matter content, the $\beta$-function in each node vanishes and eight supercharges are conserved. Thus, the $\mathbb{Z}_2$ orbifold projection yields a 
$\mathcal{N}=2$ superconformal theory in which the gauge group is $\mathrm{SU}(N)\times\mathrm{SU}(N)$ and the $R$-symmetry group is 
$\mathrm{SU}(2)_R\times\mathrm{U}(1)_R$. Although one could assign a different gauge coupling $g_I$ to each node, we will focus only on the symmetric configuration in which $g_0=g_1\equiv g$. In this way we have a single 't Hooft coupling 
\begin{align}
\lambda = N g^2
\label{lambda}
\end{align}
and the planar limit is $N\to\infty$ with fixed $\lambda$.

Denoting by $\phi_{I=0,1}$ the complex scalar fields of the vector multiplets in the two nodes, we introduce 
for any integer $k\geq 2$ the following local chiral 
operators
\begin{align}
U_k(x) = \frac{1}{\sqrt{2}} \Big[
\textrm{tr}\,\phi_0(x)^k+\textrm{tr}\,\phi_1(x)^k \Big]~,
\quad
T_k(x) = \frac{1}{\sqrt{2}} \Big[
\textrm{tr}\,\phi_0(x)^k-\textrm{tr}\,\phi_1(x)^k \Big]~,
\label{UToperators}
\end{align}
which are conformal primaries of charge $k$ and 
dimension
\begin{align}
\Delta_{U_k} = \Delta_{T_k} = k ~.    
\end{align}
The operators $U_k(x)$ and $T_k(x)$ are called, respectively, 
untwisted and twisted\,%
\footnote{The holographic origin of this terminology has been discussed in \cite{Billo:2022fnb}.}, since under the exchange $\phi_0 \longleftrightarrow \phi_{1}$ which generates the $\mathbb{Z}_2$ group they are even and odd and thus correspond to the untwisted and twisted sectors of the $\mathbb{Z}_2$ orbifold.
The anti-chiral operators $\overbar{U}_k(x)$ and $\overbar{T}_k(x)$ are obtained by replacing the complex scalar fields $\phi_{0,1}$ with their complex conjugates $\overbar{\phi}_{0,1}$, so that we simply have 
$\overbar{U}_k(x)=[U_k(x)]^*$ and $\overbar{T}_k(x)=[T_k(x)]^*$.

The quantities we are interested in are the 2- and 3-point functions 
among the operators \eqref{UToperators} and their conjugates. These correlators are constrained by conformal invariance, the conservation of the $U(1)_R$-charge and the symmetries of the $\mathbb{Z}_2$ quiver, which require that the sum of the conformal dimensions of the chiral operators must match that of the anti-chiral ones and that the number of twisted operators must be even in order to have a non-vanishing result. These constraints fix the form of the non-zero 2-point functions to be
\begin{align}
\label{eq:2point}
    \big\langle U_k(x) \, \overbar{U}_k(y) \big\rangle = \frac{G_{U_k}}{|x-y|^{2k}}
    ~, \quad \big\langle T_k(x) \, \overbar{T}_k(y) \big\rangle = \frac{G_{T_k}}{|x-y|^{2k}} ~,
\end{align}
where the coefficients $G_{U_k}$ and $G_{T_k}$ are functions of $N$ and 
$\lambda$. Likewise, the only non-vanishing 3-point functions
are\,%
\footnote{Of course, there are also the conjugate correlators in which the chiral/anti-chiral operators are exchanged.} 
\begin{subequations}
\begin{align}
& \big\langle U_k(x) \, U_\ell(y) \, \overbar{U}_{p}(z) \big\rangle = \frac{G_{U_k,U_\ell,\overbar{U}_{p}}}{|x-z|^{2k}\,|y-z|^{2\ell}}~, \label{3pointUUU} \\
& \big\langle U_k(x) \, T_{\ell}(y) \, \overbar{T}_p(z) \big\rangle = \frac{G_{U_k,T_{\ell},\overbar{T}_p}}{|x-z|^{2k}\,|y-z|^{2\ell}} ~, \label{3pointUTT} \\
& \big\langle T_k(x) \, T_{\ell}(y) \, \overbar{U}_p(z) \big\rangle = \frac{G_{T_k,T_{\ell},\overbar{U}_p}}{|x-z|^{2k}\,|y-z|^{2\ell}} ~, \label{3pointTTUbar}
\end{align}
\label{3point}%
\end{subequations}
with the understanding that $p=k+\ell$ as required by the conservation of the U$(1)_R$ charge. The coefficients appearing in the numerators of \eqref{3point} are functions of $N$ and $\lambda$.

The above 2- and 3-point functions clearly depend on the normalization of the operators \eqref{UToperators}. To remove this dependence we consider the so-called
structure constants defined as
\begin{align}
& C_{U_k,U_{\ell},\overbar{U}_p} = \frac{G_{U_k,U_{\ell},\overbar{U}_p}}{\sqrt{G_{U_k}G_{U_{\ell}}G_{U_p}}}~, \quad
C_{U_k,T_{\ell},\overbar{T}_p} = \frac{G_{U_k,T_{\ell},\overbar{T}_p}}{\sqrt{G_{U_k}G_{T_{\ell}}G_{T_p}}}~, \quad
C_{T_k, T_{\ell},\overbar{U}_p} = \frac{G_{T_k,T_{\ell},\overbar{U}_p}}{\sqrt{G_{T_k} G_{T_{\ell}}G_{U_p}}} ~. \label{eq:C} 
\end{align}
As discussed in \cite{Billo:2022gmq,Billo:2022fnb}, in the planar limit the 2- and 3-point functions involving only untwisted operators do not depend on the 't Hooft coupling but only on $N$. Indeed one finds\,%
\footnote{From now on, all formulas we write are meant to be valid at the leading order for $N\to\infty$.}
\begin{subequations}
\begin{align}
G_{U_k} &= k\,\Big(\frac{N}{2}\Big)^{k}\equiv \mathcal{G}_k~,
\label{Gkuntw}\\[1mm]
G_{U_k,U_{\ell},\overbar{U}_{p}} &= \frac{k \, \ell \, p}{2\sqrt{2}}
\,\Big(\frac{N}{2}\Big)^{\frac{k+\ell+p}{2}-1} \equiv \mathcal{G}_{k,\ell,p}~.
\label{Gklpuntw}
\end{align}\label{Guntwisted}%
\end{subequations}
Thus, also the untwisted structure constants are $\lambda$ independent and read
\begin{equation}
C_{U_k,U_{\ell},\overbar{U}_{p}}=\frac{1}{\sqrt{2}\,N}\,\sqrt{k\,\ell\,p}~.
\label{Cuntwisted}
\end{equation}
Apart from the factor of $\sqrt{2}$ due to the orbifold, this is the same expression as in $\mathcal{N}=4$ SYM \cite{Lee:1998bxa}. Instead, the coefficients in the correlators involving twisted operators and the corresponding structure constants depend on $\lambda$ in a nontrivial way, and our goal in the next sections is to find their exact expression and their expansion at large $\lambda$.

\section{Localization}
\label{sec:localization}

The gauge-theory correlators \eqref{eq:2point} and \eqref{3point} are defined in flat space but can be conformally mapped to analogous correlators on a 4-sphere $S^4$. Through this map it is possible to evaluate the coefficients $G$ by exploiting the power of supersymmetric localization \cite{Pestun:2007rz} with matrix-model techniques as discussed in \cite{Baggio:2014sna,Baggio:2015vxa,Gerchkovitz:2016gxx,Baggio:2016skg,Rodriguez-Gomez:2016ijh,Rodriguez-Gomez:2016cem,Pini:2017ouj,Billo:2017glv,Billo:2019fbi,Beccaria:2020azj,Beccaria:2020hgy,Galvagno:2020cgq,Beccaria:2021hvt,Fiol:2021icm,Billo:2021rdb,Billo:2022xas,Billo:2022gmq,Billo:2022fnb}.

\subsection{The matrix model}
For the quiver theory of Fig.~\ref{fig:quiver}, the matrix model consists of two $N \times N$ Hermitian traceless matrices $a_{I}$ taking values in the $\mathfrak{su}(N)$ Lie-algebra of the $I$-th gauge group, namely
\begin{align}
    a_{I} = a_I^b\,T_b, \ \ \ \ \  b=1,\dots,N^2-1 ~.
\end{align}
Here $T_b$ are the generators of the $\mathfrak{su}(N)$ Lie-algebra in the fundamental representation, normalized so that
\begin{align}
    \textrm{tr} \,T_b\,T_c = \frac{1}{2}\,\delta_{b,c} \ \ \ \ \ \ \  b,c=1,\dots,N^2-1 ~.
\end{align}
In the large-$N$ limit, we can neglect the non-perturbative instanton contributions that are exponentially suppressed and, adopting the so-called full Lie-algebra approach \cite{Billo:2017glv}, we can write the partition function of the matrix model as
\begin{align}
    \mathcal{Z} = \int \!da_0 \, da_1 \ \textrm{e}^{-\textrm{tr} \, a_0^2-\textrm{tr} \, a_1^2 -S_{\mathrm{int}}} 
    = \big\langle \,\textrm{e}^{-S_{\mathrm{int}}}\, \big\rangle_{0}~.
    \label{eq:Z}
\end{align}
Here the integration measure is defined as
\begin{align}
    da_I = \prod_{b=1}^{N^2-1}\frac{da_I^b}{\sqrt{2\pi}}~,
\end{align}
and the interaction action $S_{\mathrm{int}}$ is \cite{Billo:2021rdb}
\begin{align}
\label{Sint}
S_{\mathrm{int}} = 
2\sum_{m=2}^{\infty}\sum_{k=2}^{2m}(-1)^{m+k}\Big(\frac{\lambda}{8\pi^2N}\Big)^{m}\,\binom{2m}{k}\,
\frac{\zeta_{2m-1}}{2m}\,\big(\textrm{tr} \, a_0^{2m-k}-\textrm{tr}\,a_{1}^{2m-k}\big) \big(\textrm{tr}\, a_0^k - \textrm{tr} \, a_{1}^k\big)     
\end{align}
where $\zeta_{2m-1}$ are the Riemann $\zeta$-values. Finally, in (\ref{eq:Z}) the notation $\langle\,\cdot\,\rangle_0$ stands for the expectation value in the free Gaussian model.

In analogy with the gauge-theory operators (\ref{UToperators}), in the matrix model we introduce the following untwisted and twisted combinations
\begin{align}
\label{A}
A_{k}^{\pm} = \frac{1}{\sqrt{2}}\big(\textrm{tr} \, a_0^k \pm \textrm{tr} \, a_1^k \big)~,
\end{align}
for $k \geq 2$. By normal ordering them, we obtain the matrix-model representatives of
the operators (\ref{UToperators}) and can establish the following correspondence
\begin{align}
\label{correspondence}
    U_k(x) \longleftrightarrow \,\,:\!A_{k}^{+}\!: \,\,\equiv\,{O}_{k}^{+}
     \qquad \textrm{and} \qquad  T_k(x) \longleftrightarrow \,\,:\!A_{k}^{-}\!: \,\,\equiv\,{O}_{k}^{-}~.
\end{align}
Here the symbol $:\,:$ denotes, as usual, the normal ordering, namely the subtraction of the contractions
of $A_k^{\pm}$ with all single-trace operators of lower dimension. As shown in \cite{Baggio:2016skg,Billo:2022fnb}, in the planar limit this subtraction is equivalent to perform a Gram-Schmidt orthogonalization within the set of single-trace operators only. Thus we can write
\begin{align}
    O_k^\pm=\sum_{\ell\leq k}\mathsf{M}_{k,\ell}^\pm\,A_\ell^{\pm}
    \label{OAk}
\end{align}
where the mixing coefficients $\mathsf{M}_{k,\ell}^\pm$ are determined by requiring that $O_k^\pm$ be orthogonal to all operators with lower dimensions. The twisted sector coefficients $\mathsf{M}_{k,\ell}^-$ depend on $\lambda$ and $N$, while the untwisted coefficients $\mathsf{M}_{k,\ell}^+$ are functions only of $N$.
Notice also that the anti-chiral operators $\overbar{U}_k(x)$ and $\overbar{T}_k(x)$ are represented in the matrix model by the same operators ${O}^{\pm}_k$ as the chiral ones.

Using the dictionary (\ref{correspondence}), we can translate the evaluation of the normalization of the correlation functions \eqref{eq:2point} and (\ref{3point}) in a matrix model computation. For example we have
\begin{align}
G_{T_k}&=\big\langle {O}^-_k \,{O}^-_k\big\rangle
=\frac{\big\langle {O}^-_k\,{O}^-_k\,\mathrm{e}^{-S_{\mathrm{int}}}\big\rangle_0\phantom{\Big|}}{{\big\langle
\mathrm{e}^{-S_{\mathrm{int}}}\big\rangle_0\phantom{\Big|}}}~,
\label{GTk}
\end{align}
and
\begin{align}
G_{U_k,T_\ell,\overbar{T}_p}&=\big\langle {O}^+_k\,
{O}^-_\ell\,{O}^-_p\big\rangle
=\frac{\big\langle {O}^+_k\,{O}^-_\ell\,
{O}^-_p\,\mathrm{e}^{-S_{\mathrm{int}}}\big\rangle_0\phantom{\Big|}}{{\big\langle
\mathrm{e}^{-S_{\mathrm{int}}}\big\rangle_0\phantom{\Big|}}}~.
\label{GUTT}
\end{align}
Therefore, in this approach everything is reduced to calculating vacuum expectation values in a two-matrix Gaussian model.

\subsection{The \texorpdfstring{$\mathsf{X}$}{} and \texorpdfstring{$\mathsf{D}$}{} matrices and their relation to the 2- and 3-point functions}
The interaction action \eqref{Sint} depends only on the twisted combinations $A^{-}_{k}$ and can be written as
\begin{align}
\label{Sint2}
S_{\mathrm{int}} = 
2\sum_{m=2}^{\infty}\sum_{k=2}^{2m}(-1)^{m+k}\Big(\frac{\lambda}{8\pi^2N}\Big)^{m}\,\binom{2m}{k}\,
\frac{\zeta_{2m-1}}{m}\,A_{2m-k}^-\,A_k^-~.
\end{align}
As shown in \cite{Beccaria:2021hvt,Billo:2021rdb}, this fact allows to express the partition function (\ref{eq:Z}) as
\begin{align}
\mathcal{Z} = \big[\det(\mathbb{1}-\mathsf{X})\big]^{-\frac{1}{2}}  ~,  
\end{align}
where $\mathsf{X}$ is a $\lambda$-dependent infinite symmetric matrix whose entries with opposite parity vanish, namely
\begin{align}
\mathsf{X}_{2n,2m+1} = 0 
\label{Xzero}
\end{align}
for $n,m\geq 1$, while those with the same parity are given by the following convolution of Bessel functions of the first kind:
\begin{align}
\label{Xmatrix}
\mathsf{X}_{k,\ell} = -8(-1)^{\frac{k+\ell+2k\ell}{2}}\sqrt{k\,\ell}\int_{0}^{\infty}\!\frac{dt}{t}\frac{e^{t}}{(e^t-1)^2}\ J_k\Big(\frac{t\sqrt{\lambda}}{2\pi}\Big)J_{\ell}\Big(\frac{t\sqrt{\lambda}}{2\pi}\Big)~,
\end{align}
for $k,\ell \geq 2$. Since the 't Hooft coupling only enters the arguments of the Bessel functions, one can use the asymptotic expansion of the latter to study the strong coupling limit of $\mathsf{X}$ \cite{Beccaria:2021hvt,Billo:2021rdb}. 
Because of (\ref{Xzero}), it is convenient to define the matrices $\mathsf{X}^{\mathrm{even}}$ and $\mathsf{X}^{\mathrm{odd}}$ according to
\begin{align}
(\mathsf{X}^{\textrm{even}})_{n,m} 
= \mathsf{X}_{2n,2m} \qquad \textrm{and} \qquad 
(\mathsf{X}^{\textrm{odd}})_{n,m} = \mathsf{X}_{2n+1,2m+1}
\label{Xevenodd}
\end{align}
for $n,m\geq 1$. 

In \cite{Beccaria:2020hgy} it has been shown that for any value of $\lambda$ the 2-point functions $G_{T_k}$ can be expressed in a closed compact form as\,%
\footnote{With an abuse of notation, here and in the following we denote by $\frac{1}{\mathsf{M}}$ the inverse of the matrix $\mathsf{M}$.}
\begin{equation}
G_{T_{2n}}=\mathcal{G}_{2n}\,\Big(\frac{1}{\mathbb{1}-\mathsf{X}^{\mathrm{even}}_{[n]}}\Big)_{1,1}
\qquad\mbox{and}\qquad
G_{T_{2n+1}}=\mathcal{G}_{2n+1}\,\Big(\frac{1}{\mathbb{1}-\mathsf{X}^{\mathrm{odd}}_{[n]}}\Big)_{1,1}
\label{GTbis}
\end{equation}
where $\mathcal{G}_k$ are the $\mathcal{N}=4$ coefficients 
defined in (\ref{Gkuntw}), while $\mathsf{X}^{\mathrm{even}}_{[n]}$ and $\mathsf{X}^{\mathrm{odd}}_{[n]}$ are the sub-matrices obtained from $\mathsf{X}^{\mathrm{even}}$ and $\mathsf{X}^{\mathrm{odd}}$
by removing their first $(n-1)$ rows and columns\,%
\footnote{Clearly one has
$\mathsf{X}_{[1]}^{\mathrm{even}}=\mathsf{X}^{\mathrm{even}}$ and $\mathsf{X}_{[1]}^{\mathrm{odd}}=\mathsf{X}^{\mathrm{odd}}$.}.
Using Cramer's rule, we can deduce from (\ref{GTbis}) that
\begin{align}
R_{2n}\,\equiv\, \frac{G_{T_{2n}}}{\mathcal{G}_{2n}}=\frac{\det\big(\mathbb{1}-\mathsf{X}^{\mathrm{even}}_{[n+1]}\big)}{\det\big(\mathbb{1}-\mathsf{X}^{\mathrm{even}}_{[n]}\big)}
\qquad\mbox{and}\qquad
R_{2n+1}\,\equiv\, \frac{G_{T_{2n+1}}}{\mathcal{G}_{2n+1}}=\frac{\det\big(\mathbb{1}-\mathsf{X}^{\mathrm{odd}}_{[n+1]}\big)}{\det\big(\mathbb{1}-\mathsf{X}^{\mathrm{odd}}_{[n]}\big)}~.
\label{GTter}
\end{align}
In this way the calculation of $G_{T_k}$ is reduced to the calculation of the determinants of the matrices $\big(\mathbb{1}-\mathsf{X}^{\mathrm{even}}_{[n]}\big)$
and $\big(\mathbb{1}-\mathsf{X}^{\mathrm{odd}}_{[n]}\big)$ for different values of $n$. These determinants have the form of Fredholm determinants of Bessel operators \cite{Beccaria:2022ypy}.

The ratios $R_{2n}$ and $R_{2n+1}$ can be written in yet another way by introducing the symmetric matrix
\begin{equation}
\mathsf{D}=\frac{1}{\mathbb{1}-\mathsf{X}}=\mathbb{1}+\mathsf{X}+\mathsf{X}^2+\mathsf{X}^3+\dots
\label{D}
\end{equation}
whose elements $\mathsf{D}_{k,\ell}$ (with $k,\ell\geq 2$) vanish unless $k$ and $\ell$ have the same parity
as a consequence of (\ref{Xzero}). Indeed, defining the matrices $\mathsf{D}^{\mathrm{even}}$ and 
$\mathsf{D}^{\mathrm{odd}}$ in analogy with (\ref{Xevenodd}), one can show \cite{Beccaria:2020hgy} that
\begin{align}
R_{2n}=\frac{\det\big(\mathsf{D}^{\mathrm{even}}_{(n)}\big)}{\det\big(\mathsf{D}^{\mathrm{even}}_{(n-1)}\big)}
\qquad\mbox{and}\qquad
R_{2n+1}=\frac{\det\big(\mathsf{D}^{\mathrm{odd}}_{(n)}\big)}{\det\big(\mathsf{D}^{\mathrm{odd}}_{(n-1)}\big)}
\label{GTquater}
\end{align}
where $\mathsf{D}^{\mathrm{even}}_{(n)}$ and $\mathsf{D}^{\mathrm{odd}}_{(n)}$ denote, respectively, 
the upper-left $n\times n$ blocks of $\mathsf{D}^{\mathrm{even}}$ and $\mathsf{D}^{\mathrm{odd}}$, with the understanding that $\mathsf{D}^{\mathrm{even}}_{(0)}=\mathsf{D}^{\mathrm{odd}}_{(0)}=\mathbb{1}$. 
For example we have
\begin{equation}
    R_2=\mathsf{D}_{2,2}~,\qquad R_3=\mathsf{D}_{3,3}~,\qquad
    R_4=\frac{\mathsf{D}_{2,2}\,\mathsf{D}_{4,4}-\mathsf{D}_{2,4}^2}{\mathsf{D}_{2,2}}
    \label{Rkexp}
\end{equation}
and so on.

Let us now turn to the 3-point functions. To study them, following the approach developed in \cite{Billo:2022fnb}, we introduce the operators
\begin{align}
    \mathcal{P}_k^{\pm}=\frac{1}{\sqrt{\mathcal{G}_k}}\,O_k^\pm\Big|_{\lambda=0}
    \label{P}
\end{align}
which are normal ordered with respect to the free theory. In the untwisted case, this amounts to a simple rescaling; indeed
\begin{equation}
    O_k^+=O_k^+\Big|_{\lambda=0}=\sqrt{\mathcal{G}_k}\,\mathcal{P}_k^+
    \label{OP+}
\end{equation}
since in this sector there are no $\lambda$-dependent effects and the normal ordering in the interacting theory coincides with the normal ordering in the free theory. Instead, in the twisted sector where the interactions play a role we have
\begin{align}
    O_k^-=\sqrt{\mathcal{G}_k}\Big(\mathcal{P}_k^--\sum_{\ell<k}
    \mathsf{Q}_{k,\ell}\,\mathcal{P}_\ell^-\Big)
    \label{OP-}
\end{align}
where $\mathsf{Q}_{k,\ell}$ are $\lambda$-dependent mixing coefficients which can be computed explicitly \cite{Billo:2022fnb}. However, for what follows their expression is not needed. 

The big advantage of using $\mathcal{P}_k^\pm$ instead of $A_k^\pm$ as a basis to express the operators $O_k^\pm$ consists in the fact that the 2-functions of $\mathcal{P}_k^\pm$ are simple and, most importantly, that their 3-point functions are factorized. Indeed, as shown in \cite{Billo:2022fnb}, one has
\begin{align}
\big\langle \mathcal{P}_k^+ \,\mathcal{P}_\ell^+\big\rangle = \delta_{k,\ell}~,\qquad
\big\langle \mathcal{P}_k^- \,\mathcal{P}_\ell^-\big\rangle = \mathsf{D}_{k,\ell}~,\label{Dkl}
\end{align}
and 
\begin{subequations}
\begin{align}
\big\langle \mathcal{P}_k^+\,\mathcal{P}_\ell^+\,\mathcal{P}_p^+
\big\rangle&=\frac{1}{\sqrt{2}\,N}\,\sqrt{k\,\ell\,p}~,\\
\big\langle \mathcal{P}_k^+\,\mathcal{P}_\ell^-\,\mathcal{P}_p^-
\big\rangle&=\frac{1}{\sqrt{2}\,N}\,\sqrt{k}\,\,\mathsf{d}_\ell\,
\mathsf{d}_p~,\\
\big\langle \mathcal{P}_k^-\,\mathcal{P}_\ell^-\,\mathcal{P}_p^+
\big\rangle&=\frac{1}{\sqrt{2}\,N}\,\mathsf{d}_k\,
\mathsf{d}_\ell\,\sqrt{p}~,
\end{align}%
\end{subequations}
where $p=k+\ell$ and
\begin{equation}
    \mathsf{d}_k=\sum_{k^\prime}\mathsf{D}_{k,k^\prime}\,\sqrt{k^\prime}
    ~.
    \label{dk}
\end{equation}
Therefore, using (\ref{OP+}), (\ref{OP-}) and the above 3-point functions, one immediately finds:
\begin{subequations}
\begin{align}
G_{U_k,U_\ell,\overbar{U}_p}&=\big\langle {O}_k^+\,{O}_\ell^+\,{O}_p^+
\big\rangle=\frac{\sqrt{\mathcal{G}_k\,\mathcal{G}_\ell\,
\mathcal{G}_p}}{\sqrt{2}\,N}\,\sqrt{k\,\ell\,p} \,=\, \mathcal{G}_{k,\ell,p}~,\label{GUUUklp}\\[1mm]
G_{U_k,T_\ell,\overbar{T}_p}&=\big\langle {O}_k^+\,{O}_\ell^-\,{O}_p^-
\big\rangle=\frac{\sqrt{\mathcal{G}_k\,\mathcal{G}_\ell\,
\mathcal{G}_p}}{\sqrt{2}\,N}\,\sqrt{k}\,\,\widetilde{\mathsf{d}}_\ell\,
\widetilde{\mathsf{d}}_p\,=\,
\frac{\mathcal{G}_{k,\ell,p}}{\sqrt{\ell\,p}}\,
\,\widetilde{\mathsf{d}}_\ell\,\widetilde{\mathsf{d}}_p~,\label{GUTTklp}\\[1mm]
G_{T_k,T_\ell,\overbar{U}_p}&=\big\langle {O}_k^-\,{O}_\ell^-\,{O}_p^+
\big\rangle=\frac{\sqrt{\mathcal{G}_k\,\mathcal{G}_\ell\,
\mathcal{G}_p}}{\sqrt{2}\,N}\,\widetilde{\mathsf{d}}_k\,\widetilde{\mathsf{d}}_\ell\,\sqrt{p}\,=\,
\frac{\mathcal{G}_{k,\ell,p}}{\sqrt{k\,\ell}}\,
\,\widetilde{\mathsf{d}}_k\,\widetilde{\mathsf{d}}_\ell~,
\label{GTTUklp}
\end{align}
\label{G3klp}%
\end{subequations}
where $p=k+\ell$ and
\begin{equation}
    \widetilde{\mathsf{d}}_k=\mathsf{d}_k-\sum_{\ell<k}\mathsf{Q}_{k,\ell}\,\mathsf{d}_\ell~.
\label{tildedk}
\end{equation}
In other words the 3-point functions of the $O_k$ operators have exactly the same form of those of the $\mathcal{P}_k$ operators but with $\mathsf{d}_k$ replaced by $\widetilde{\mathsf{d}}_k$ and a different overall normalization.

Eq.~(\ref{GUUUklp}) confirms the expected result (\ref{Guntwisted}), while (\ref{GUTTklp}) and (\ref{GTTUklp}) provide a simple expression of the 3-point functions with twisted operators in terms of the $\lambda$-dependent quantities $\widetilde{\mathsf{d}}_k$. Since also the mixing coefficients $\mathsf{Q}_{k,\ell}$ are functions of $\mathsf{D}_{k,\ell}$, ultimately everything can be written in terms of such matrix elements.
The explicit form of these expressions can be worked out on a case-by-case basis without particular difficulties, but we have not found a simple closed-form result. Quite remarkably, however, we can bypass this problem since, as we will show in the next sections, it is possible to link the combination $\widetilde{\mathsf{d}}_k$ in (\ref{tildedk}) directly to the ratios $R_k$ defined in (\ref{GTter}) through a simple exact relation valid for all values of $\lambda$.

\section{The 2-point functions at strong coupling}
\label{sec:2point}

From the asymptotic expansion of the Bessel functions, one can derive the behavior of the $\mathsf{X}$ and $\mathsf{D}$ matrices when the 't Hooft coupling becomes large. As shown in \cite{Beccaria:2021hvt,Billo:2022xas,Billo:2022fnb} at leading order one finds
\begin{align}
& \mathsf{D}_{k,\ell} \ \underset{\lambda \rightarrow \infty}{\sim} \ \frac{\pi^2}{\lambda}\Big[\sqrt{k\,\ell} \ (\textrm{min}(k^2,\ell^2)
-\delta_{k \,\textrm{mod} \, 2,1})\Big] +O\Big(\frac{1}{\lambda^{3/2}}\Big)
\label{Dstrong}
\end{align}
for any $k,\ell\geq 2$. Using this result in (\ref{GTquater}) one easily obtains the
leading strong-coupling behavior of the 2-point functions, namely
\cite{Billo:2022gmq,Billo:2022fnb}
\begin{align}
G_{T_k}^{(\mathrm{LO})}=\mathcal{G}_k\,\frac{4\pi^2}{\lambda}\,k(k-1)
\end{align}
for $k\geq 2$. 

The full strong-coupling expansion of $G_{T_k}$ has been recently derived in \cite{Beccaria:2022ypy} using the same techniques that have been previously used to study the octagon form-factors in $\mathcal{N}=4$ SYM \cite{Belitsky:2019fan,Belitsky:2020qrm,Belitsky:2020qir}. More precisely, observing that $\det\big(\mathbb{1}-\mathsf{X}^{\mathrm{even}}_{[n]}\big)$ and $\det\big(\mathbb{1}-\mathsf{X}^{\mathrm{odd}}_{[n]}\big)$
can be interpreted as Fredholm determinants of Bessel operators, the authors of \cite{Beccaria:2022ypy}
have obtained\,%
\footnote{Here and in the following we consider only the ``perturbative'' parts of the strong-coupling expansions that contain terms proportional to inverse powers of the 't Hooft coupling. As noticed in \cite{Beccaria:2022ypy}, these ``perturbative'' strong-coupling series are non Borel summable due to the existence of singularities on the positive real axis of the Borel plane, and must be completed with series of non-perturbative terms proportional to powers of $\mathrm{e}^{-\sqrt{\lambda}}$.
\label{footnote_np}}
\begin{equation}
\begin{aligned}
\log \Big[ \det \big(\mathbb{1}-\mathsf{X}^{\mathrm{even}}_{[n]}\big) \Big]&= \frac{\sqrt{\lambda}}{4} -\Big(2n-\frac{3}{2}\Big)\log\Big(\frac{\sqrt{\lambda}}{4\pi}\Big) +B_{2n-1}+ f_{2n-1}~,\\
\log \Big[ \det \big(\mathbb{1}-\mathsf{X}^{\mathrm{odd}}_{[n]}\big) \Big]&= \frac{\sqrt{\lambda}}{4} -\Big(2n-\frac{1}{2}\Big)\log\Big(\frac{\sqrt{\lambda}}{4\pi}\Big) +B_{2n}+ f_{2n}~,
\end{aligned}
\label{logdetX}
\end{equation}
where
\begin{align}
B_k=-6 \log \mathsf{A} +\frac{1}{2} + \frac{1}{6}\log 2 -k\,\log 2 +
\log \Gamma(k)
\end{align}
with $\mathsf{A}$ being the Glaisher constant, and
\begin{align}
f_k&= \frac{1}{16}(2k-3)(2k-1)
\log\Big(\frac{\lambda^\prime}{\lambda}\Big) 
+ (2k-5)(2k-3)(4k^2-1)\frac{\zeta_3}{32\lambda^{\prime\, 3/2}} \nonumber \\
& ~~- (2k-7)(2k-5)(4k^2-9)(4k^2-1)\frac{3\,\zeta_5}{256 \lambda^{\prime\,5/2}} \nonumber \\
&~~ - (2k-5)(2k-3)(4k^2-1)(4k^2-8k-17)\frac{3\,\zeta_3^2}{64
\lambda^{\prime\,3}} + O\Big(\frac{1}{\lambda^{\prime\,7/2}}\Big)+\cdots~.
\end{align}
Here and in the following, the ellipses indicate non-perturbative exponentially small corrections mentioned in footnote~\ref{footnote_np}, and
\begin{equation}
\sqrt{\lambda^\prime} = \sqrt{\lambda} -4\log 2~.
\label{lambdaprime}
\end{equation}
Of course, the sub-leading terms in $1/\lambda^\prime$ can be computed to any desired order.
Inserting these expansions into (\ref{GTter}), after some simple manipulations one finds
\begin{align}
G_{T_k}=G_{T_k}^{(\mathrm{LO})}\ &
\Big(\frac{\lambda^\prime}{\lambda}\Big)^{k-1}\bigg[1+(k-1)(2k-1)(2k-3)\,\frac{\zeta_3}{\lambda^{\prime\,3/2}}\notag\\
&~-(k-1)(2k-3)(2k-5)(4k^2-1)\,\frac{9\,\zeta_5}{16\lambda^{\prime\,5/2}}\notag\\
&~+(k-1)(2k-1)(2k-3)(2k-5)(4k^2-20k-3)\,\frac{\zeta_3^2}{4\lambda^{\prime\,3}}+O\Big(\frac{1}{\lambda^{\prime\,7/2}}\Big)\bigg]+\cdots
\label{GTkexpanded}
\end{align}
for any $k\geq 2$. Expressing $\lambda^\prime$ in terms of $\lambda$ by means of (\ref{lambdaprime}) and expanding for large $\lambda$, one obtains the expansion in terms of the original t' Hooft coupling. One can easily check that the first two contributions beyond leading order in this expansion match those computed in \cite{Bobev:2022grf} with high-precision numerical methods.

\section{A special class of 3-point functions}
\label{sec:diagonal}

Before addressing the strong-coupling expansion of the 3-point functions in full generality, we consider the cases of the form
\begin{align}
    G_{T_k,T_k,\overbar{U}_{2k}}=
    \big\langle {O}^-_k\,{O}^-_k \,{O}^+_{2k}\big \rangle~.
    \label{GTkTkU2k}
\end{align}
As we are going to see, these special 3-point functions are related to the 2-point functions $G_{T_k}$ studied in the previous section by means of a Ward-like identity which is valid for all $\lambda$. Through this identity, the strong-coupling expansion of $G_{T_k,T_k,\overbar{U}_{2k}}$ can be directly obtained from (\ref{GTkexpanded}).

To derive this result, we use the basis of operators $A_k^{\pm}$ defined in (\ref{A}) and introduce the convenient notation
\begin{align}
    T^{\pm,\pm,\dots}_{n_{1},n_{2},\dots}= \big\langle A^{\pm}_{n_1} A^{\pm}_{n_{2}}\dots \big\rangle
    \label{notation}
\end{align}
for their vacuum expectation values. According to (\ref{OAk}), the operators appearing in the correlator (\ref{GTkTkU2k}) are:\,%
\footnote{Notice that, following the same approach discussed in \cite{Billo:2022xas} for the $\mathbf{E}$ theory, in the untwisted operators (\ref{MdefU}) we have subtracted from $A_{2p}^+$ the vacuum expectation value $T_{2p}^+$. This automatically ensures that $\langle
\mathcal{O}_{2p}^+\rangle=0$.}
\begin{subequations}
\begin{align}
\label{MdefT}
O_k^-&=\sum_{n=2}^k \mathsf{M}_{k,n}^-\,A_\ell^-~,\\
\label{MdefU}
O_{2k}^+&=\sum_{p=1}^k \mathsf{M}_{2k,2p}^+\,\big(A_{2p}^+-T_{2p}^+\big)~.
\end{align}
\label{Mdef}%
\end{subequations}
The mixing coefficients $\mathsf{M}_{k,n}^-$ and $\mathsf{M}_{2k,2p}^+$ are determined with the Gram-Schmidt procedure by requiring that $O_k^-$ and $O_{2k}^+$ be orthogonal to all operators of lower conformal dimensions. For the untwisted operators, 
these coefficients do not depend on the t' Hooft coupling and thus can be computed in the free theory. They are given by (see for example Eq.\,(5.9) in \cite{Billo:2022xas})
\begin{align}
\mathsf{M}_{2k,2p}^+=\Big(\!-\frac{N}{2}\Big)^{k-p}\,\,\frac{k}{p}\,
\binom{k+p-1}{k-p}~.
\label{M+}
\end{align}
For the twisted operators, instead, the coefficients $\mathsf{M}_{k,n}^-$ are nontrivial functions of $\lambda$; however, fortunately, for the current calculation we do not need their explicit expression.

Using \eqref{Mdef}, the 3-point function (\ref{GTkTkU2k}) becomes
\begin{align}
\label{2kkk}
G_{T_k,T_k,\overbar{U}_{2k}}=\sum_{n=2}^{k}\sum_{m=2}^k
\sum_{p=1}^k
\mathsf{M}^-_{k,n}\,\mathsf{M}^-_{k,m}\,\mathsf{M}^+_{2k,2p} \,
\Big( T^{+,-,-}_{2p,n,m}-T^+_{2p}\,T^{-,-}_{n,m} \Big)~.
\end{align}
The combination of expectation values appearing in brackets can evaluated
in the planar limit using the $\mathbb{Z}_2$-quiver generalization of the relation presented in Eq.\,(B.26) of \cite{Billo:2022xas} for the $\mathbf{E}$ theory. In the quiver case this relation reads
\begin{align}
\label{identityA}
    T^{+,-,-}_{2p,n,m}-T^+_{2p}\,T^{-,-}_{n,m} 
    &= T_{2p}^+\,\frac{p(p+1)}{4N^2} \big(n+m+2\lambda\partial_{\lambda} \big)T^{-,-}_{n,m}\notag\\
    &=\frac{\sqrt{2}\,N^{p+1}}{2^p}\frac{(2p)!}{p!(p+1)!}
    \frac{p(p+1)}{4N^2}\big(n+m+2\lambda\partial_{\lambda} \big)T^{-,-}_{n,m}~,
\end{align}
where in the second line we have used the expression of $T_{2p}^+$ in the
planar limit.

When we insert this result in the right-hand-side of (\ref{2kkk}), we realize that the sum over $p$ can be factorized. Using the explicit expression of $M^+_{2k,2p}$ given in (\ref{M+}) we find  
\begin{align}
    \sum_{p=1}^{k} \mathsf{M}^+_{2k,2p} \,
    \frac{\sqrt{2}\,N^{p+1}}{2^p}\frac{(2p)!}{p!(p+1)!}\frac{p(p+1)}{4N^2}
    =\frac{k}{2\sqrt{2}}\,\Big(\frac{N}{2} \Big)^{k-1}~.
\end{align}
Therefore, our 3-point function becomes
\begin{align}
G_{T_k,T_k,\overbar{U}_{2k}}=\frac{k}{2\sqrt{2}}\,\Big(\frac{N}{2} \Big)^{k-1}
\sum_{n=2}^{k}\sum_{m=2}^k
\mathsf{M}^-_{k,n}\,\mathsf{M}^-_{k,m}\,
\big(n+m+2\lambda\partial_{\lambda} \big)T^{-,-}_{n,m}
\label{2kkk1}
\end{align}
In Appendix\,\ref{appendix:A} we prove that
\begin{align}
\label{id2kkk0}
    \frac{1}{2}\sum_{n=2}^{k}\sum_{m=2}^{k} \mathsf{M}^-_{k,n}\,
    \mathsf{M}^-_{k,m} \big( n+m+2\lambda\partial_{\lambda} \big) T^{-,-}_{n,m}=
    \big( k+\lambda\partial_{\lambda} \big) \big\langle {O}_k^-\,{O}^-_k \big\rangle ~.
\end{align}
Thus, we can rewrite (\ref{2kkk1}) in the form of a Ward-like relation between the 3-point function $G_{T_k,T_k,\overbar{U}_{2k}}$ and the 2-point function $G_{T_k}$, namely
\begin{align}
\label{fin2kkk}
    G_{T_k,T_k,\overbar{U}_{2k}} = \frac{k}{\sqrt{2}}\,\Big(\frac{N}{2} \Big)^{k-1}\big(
    k+\lambda\partial_{\lambda} \big) G_{T_k}~.
\end{align}
This result will be instrumental for the calculations described in the next section.

Exploiting this relation and using the strong-coupling expansion of $G_{T_k}$ given in (\ref{GTkexpanded}), we immediately find
\begin{align}
G_{T_k,T_k,\overbar{U}_{2k}}&=G_{T_k,T_k,\overbar{U}_{2k}}^{(\mathrm{LO})}
\,
\Big(\frac{\lambda^\prime}{\lambda}\Big)^{\frac{2k-3}{2}}\bigg[1+(2k-1)(2k-3)(2k-5)\,\frac{\zeta_3}{2\lambda^{\prime\,3/2}}\notag\\
&\quad-(2k-3)(2k-5)(2k-7)(4k^2-1)\,\frac{9\,\zeta_5}{32\lambda^{\prime\,5/2}}\notag\\
&\quad+(k-4)(2k-1)(2k-3)(2k-5)(4k^2-20k-3)\,\frac{\zeta_3^2}{4\lambda^{\prime\,3}}+O\Big(\frac{1}{\lambda^{\prime\,7/2}}\Big)\bigg] +\cdots
\label{G2kkkexpanded}
\end{align}
where the prefactor is the leading-order term \cite{Billo:2022gmq,Billo:2022fnb}
\begin{align}
G_{T_k,T_k,\overbar{U}_{2k}}^{(\mathrm{LO})}=
\mathcal{G}_{k,k,2k}\,\frac{4\pi^2}{\lambda}\,(k-1)^2
\label{GTkk2kLO}
\end{align}
with $\mathcal{G}_{k,k,2k}$ being the coefficients defined in (\ref{Gklpuntw}). We note that the first two terms arising from the expansion of the ratio $\lambda^\prime/\lambda$ for large $\lambda$ agree with the 
extrapolation of the numerical results reported in \cite{Bobev:2022grf}.

Combining (\ref{G2kkkexpanded}) and (\ref{GTkexpanded}), it is straightforward to find that the strong-coupling expansion of the structure constants for this particular family is
\begin{align}
C_{T_k,T_k,\overbar{U}_{2k}}=C_{T_k,T_k,\overbar{U}_{2k}}^{(\mathrm{LO})}
\,
\Big(\frac{\lambda}{\lambda^\prime}\Big)^{\frac{1}{2}}&\bigg[1-(4\delta_k^2-1)\,\frac{3\,\zeta_3}{2\lambda^{\prime\,3/2}}+(4\delta_k^2-1)(4\delta_k^2-9)\,\frac{45\,\zeta_5}{32\lambda^{\prime\,5/2}}\notag\\
&\quad+(4\delta_k^2-1)(12\delta_k^2-19)\,\frac{9\,\zeta_3^2}{4\lambda^{\prime\,3}}+O\Big(\frac{1}{\lambda^{\prime\,7/2}}\Big)\bigg] +\cdots	\label{C2kkkexpanded}
\end{align}
where
$C_{T_k,T_k,\overbar{U}_{2k}}^{(\mathrm{LO})}$ is the leading-order term \cite{Billo:2022gmq,Billo:2022fnb}
\begin{align}
C_{T_k,T_k,\overbar{U}_{2k}}^{(\mathrm{LO})}=\frac{\sqrt{2k(k-1)^2}}{\sqrt{2}\,N}
\label{C2kkkLO}
\end{align}
and the sub-leading corrections have been written in terms of the shifted conformal dimension of the twisted operators
\begin{equation}
\delta_k=k-1~.
\label{deltak}
\end{equation}
Indeed, as we have seen in \cite{Billo:2022gmq,Billo:2022fnb}, it is this shifted conformal dimension that enters in the various formulas when $\lambda\to\infty$. Since the leading term (\ref{C2kkkLO})
precisely matches the holographic result obtained using the AdS/CFT correspondence in the $\mathbb{Z}_2$
orbifold background at the supergravity level \cite{Billo:2022gmq,Billo:2022fnb}, we expect that the sub-leading corrections in the square brackets of (\ref{C2kkkexpanded}) are due to the higher-derivative string corrections to the supergravity action.

\section{The general 3-point functions at strong coupling}
\label{sec:general}
In Section~\ref{sec:localization} we have seen that the 3-point functions
with twisted operators can be expressed as products of the $\lambda$-dependent combinations $\widetilde{\mathsf{d}}_k$ defined in (\ref{tildedk}). Thus, to find their strong-coupling expansion we have to determine how $\widetilde{\mathsf{d}}_{k}$ behaves when $\lambda\to \infty$. This is possible by exploiting the results of the previous section.

To see this, let us first rewrite (\ref{fin2kkk}) introducing the ratio $R_k=G_{T_k}/\mathcal{G}_k$ and get
\begin{align}
    G_{T_k,T_k,\overbar{U}_{2k}} &= \frac{k\,\mathcal{G}_k}{\sqrt{2}}\,\Big(\frac{N}{2} \Big)^{k-1}\big(
    k+\lambda\partial_{\lambda} \big) R_{k}\notag\\ 
    &= \frac{\mathcal{G}_{k,k,2k}}{k}\,\big(
    k+\lambda\partial_{\lambda} \big) R_{k}
    \label{ward1}
\end{align}
where the second line follows from the explicit expressions of the $\mathcal{G}$ coefficients given in (\ref{Guntwisted}). Then, by comparing (\ref{ward1}) with (\ref{GTTUklp}) evaluated for $\ell=k$ and $p=2k$, we read that
\begin{equation}
\widetilde{\mathsf{d}}_k=\sqrt{\big(
    k+\lambda\partial_{\lambda} \big) R_{k}}~=~\frac{1}{\sqrt{\mathcal{G}_k}}\,\sqrt{\big(
    k+\lambda\partial_{\lambda} \big) G_{T_k}}~.
    \label{relation}
\end{equation}
This relation is valid for all values of the 't Hooft coupling and allows us to obtain the strong-coupling behavior of $\widetilde{\mathsf{d}}_k$ directly from that of the 2-point functions
without having to calculate explicitly the behavior of the mixing coefficients $\mathsf{Q}_{k,\ell}$ and the matrix elements $\mathsf{D}_{k,\ell}$.

If we use (\ref{relation}) in the general formulas (\ref{G3klp}), we obtain the announced Ward-like identities between
the 3- and the 2-point functions. Explicitly they are given by
\begin{subequations}
\begin{align}
G_{U_k,T_\ell,\overbar{T}_p}&=
\frac{\mathcal{G}_{k,\ell,p}}{\sqrt{\ell\,\mathcal{G}_\ell\,p\,\mathcal{G}_p}}
\,\,\sqrt{\big(
    \ell+\lambda\partial_{\lambda} \big) G_{T_\ell}}\,\sqrt{\big(
    p+\lambda\partial_{\lambda} \big) G_{T_p}}~,\label{GUTTward}\\[1mm]
G_{T_k,T_\ell,\overbar{U}_p}&=
\frac{\mathcal{G}_{k,\ell,p}}{\sqrt{k\,\mathcal{G}_k\,\ell\,\mathcal{G}_\ell}}
\,\,\sqrt{\big(
    k+\lambda\partial_{\lambda} \big) G_{T_k}}\,\sqrt{\big(
    \ell+\lambda\partial_{\lambda} \big) G_{T_\ell}}
\label{GTTUward}
\end{align}
\label{Gward}%
\end{subequations}
where $p=k+\ell$. Once again, we stress that these relations are valid for all values of $\lambda$. 

As an immediate consequence of the above result, it can be easily shown that the structure constants involving twisted operators are
\begin{subequations}
\begin{align}
C_{U_k,T_\ell,\overbar{T}_p}&=\frac{\sqrt{k}}{\sqrt{2}\,N}\,\sqrt{
    \ell+\lambda\partial_{\lambda}\big( \log G_{T_\ell} \big) }\,\sqrt{
    p+\lambda\partial_{\lambda}\big(\log G_{T_p} \big) }~,\label{CUTTward}\\[1mm]
C_{T_k,T_\ell,\overbar{U}_p}&=
\frac{\sqrt{p}}{\sqrt{2}\,N}\,\sqrt{
    k+\lambda\partial_{\lambda}\big( \log G_{T_k} \big) }\,\sqrt{
    \ell+\lambda\partial_{\lambda}\big(\log G_{T_\ell} \big) }
\label{CTTUward}
\end{align}
\label{Cward}%
\end{subequations}
where $p=k+\ell$.

\subsection{Strong coupling expansions}

The relations (\ref{Gward}) and (\ref{Cward}) can be used to determine the strong-coupling expansions of the 3-point functions and of the structure constants starting from the one of the 2-point functions given in Section~\ref{sec:2point}.

Inserting the expansion (\ref{GTkexpanded}) in (\ref{Gward}), one finds
\begin{align}
G_{U_k,T_\ell,\overbar{T}_{p}} 
= G_{U_k,T_\ell,\overbar{T}_{p}}^{(\mathrm{LO})}
\,\Big(\frac{\lambda^\prime}{\lambda}\Big)^{\frac{\delta_\ell+\delta_p-1}{2}} 
\bigg[1+g_{\ell,p}^{(1)}\,\frac{\zeta_3}{\lambda^{\prime\,3/2}}+g_{\ell,p}^{(2)}\,\frac{\zeta_5}{\lambda^{\prime\,5/2}}+g_{\ell,p}^{(3)}\,\frac{\zeta_3^2}{\lambda^{\prime\,3}}+O\Big(\frac{1}{\lambda^{\prime\,7/2}}\Big)\bigg] +\cdots
\label{GUTTexpanded}
\end{align}
with $p=k+\ell$.
The leading term, already computed in \cite{Billo:2022fnb}, is 
\begin{align}
G_{U_k,T_\ell,\overbar{T}_{p}}^{(\mathrm{LO})}=\mathcal{G}_{k,\ell,p}\,
\frac{4\pi ^2\,(\ell-1)\,(p-1)}{\lambda}
\end{align} 
while the sub-leading terms are a new result. The first few  expansion coefficients 
are
\begin{subequations}
\begin{align}
g_{\ell,p}^{(1)}&=\frac{1}{2}\,(\delta_\ell+\delta_p-1)\big(4\delta_\ell^2-4\delta_\ell\delta_p
+4\delta_p^2-2\delta_\ell-2\delta_p-3)~,\\
g_{\ell,p}^{(2)}&=-\frac{9}{32}\,(\delta_\ell+\delta_p-1)(16\delta_\ell^4-16\delta_\ell^3\delta_p+16\delta_\ell^2\delta_p^2-16\delta_\ell\delta_p^3+16\delta_p^4-24\delta_\ell^3\nonumber \\
 &\qquad\quad+8\delta_\ell^2\delta_p+8\delta_\ell\delta_p^2-24\delta_p^3 -64\delta_\ell^2+72\delta_\ell\delta_p-64\delta_p^2+36\delta_\ell+36\delta_p+45)~,\\
 g_{\ell,p}^{(3)}&=\frac{1}{8}\,(\delta_\ell+\delta_p-1)(16\delta_\ell^5-16\delta_\ell^4\delta_p+
 16\delta_\ell^3\delta_p^2+16\delta_\ell^2\delta_p^3-16\delta_\ell\delta_p^4+16\delta_p^5 \nonumber \\
& \qquad\quad-176\delta_\ell^4+160\delta_\ell^3\delta_p-192\delta_\ell^2\delta_p^2+160\delta_\ell\delta_p^3-176\delta_p^4+212\delta_\ell^3-60\delta_\ell^2\delta_p-60\delta_\ell\delta_p^2 \nonumber \\[1mm]
&\qquad\quad+212\delta_p^3 +512\delta_\ell^2-560\delta_\ell\delta_p+512\delta_p^2-279\delta_\ell-279\delta_p-342)
\end{align}
\label{gkl}%
\end{subequations}
where $\delta_\ell=\ell-1$ and $\delta_p=p-1$. In principle, one can push this calculation to any desired order using the expansions of the determinants (\ref{logdetX}) provided in \cite{Beccaria:2022ypy}. 

The 3-point function coefficients of the form $G_{T_k,T_\ell,\overbar{U}_p}$ can be obtained in exactly the same way and their strong-coupling expansion has the same form (\ref{GUTTexpanded}) with coefficients $g_{k,\ell}^{(i)}$
that are given again by the formulas (\ref{gkl})\,%
\footnote{As a consistency check, one can verify that the case discussed in Section~\ref{sec:diagonal} is correctly recovered by setting $k=\ell$.}. Expressing $\lambda^\prime$ in terms of $\lambda$, one can check that when the twisted operators have odd dimensions the first two coefficients in the large-$\lambda$ expansion of the 3-point functions coincide with those recently obtained in \cite{Bobev:2022grf} 
from an extrapolation of numerical results in the $\mathbf{E}$ theory.

Finally, from (\ref{Cward}) we can deduce the strong-coupling expansion of the structure constants: 
\begin{equation}
C_{U_k,T_\ell,\overbar{T}_{p}} 
= C_{U_k,T_\ell,\overbar{T}_{p}}^{(\mathrm{LO})}
\,\Big(\frac{\lambda}{\lambda^\prime}\Big)^{\frac{1}{2}} 
\bigg[1+c_{\ell,p}^{(1)}\,\frac{\zeta_3}{\lambda^{\prime\,3/2}}+c_{\ell,p}^{(2)}\,\frac{\zeta_5}{\lambda^{\prime\,5/2}}+c_{\ell,p}^{(3)}\,\frac{\zeta_3^2}{\lambda^{\prime\,3}}+O\Big(\frac{1}{\lambda^{\prime\,7/2}}\Big)\bigg] +\cdots~,
\label{CUTTfin}
\end{equation}
where $p=k+\ell$ and 
\begin{align}
C_{U_k,T_\ell,\overbar{T}_{p}}^{(\mathrm{LO})}=\frac{1}{\sqrt{2}\,N}\,\sqrt{k\,(\ell-1)(p-1)}~.
\label{CUTTlo}
\end{align}
The first expansion coefficients $c_{\ell,p}^{(i)}$ are
\begin{subequations}
\begin{align}
c_{\ell,p}^{(1)}&=-\frac{3}{2}\,(2\delta_\ell^2+2\delta_p^2-1)~,\\
c_{\ell,p}^{(2)}&=+\frac{45}{32}\,\Big[(2\delta_\ell^2+2\delta_p^2-1)(2\delta_\ell^2+2\delta_p^2
-9)+4(\delta_\ell^2-\delta_p^2)^2\Big]~,\\
 c_{\ell,p}^{(3)}&=+\frac{9}{4}\,\Big[(2\delta_\ell^2+2\delta_p^2-1)(6\delta_\ell^2+6\delta_p^2
-19)+10(\delta_\ell^2-\delta_p^2)^2\Big]
\end{align}
\label{ckl}%
\end{subequations}
with $p=k+\ell$.
The structure constants $C_{T_k,T_\ell,\overbar{U}_{p}}$ are given by the same expression (\ref{CUTTfin})
but with the role of $k$ and $p$ exchanged.

\section{Conclusions}
\label{sec:conclusions}
The main result of this work is the finding of exact relations between the 2- and 3-point functions of twisted operators in the $\mathcal{N}=2$ superconformal quiver theory
of Fig.~\ref{fig:quiver} when the number colors $N$ tends to infinity. These relations extend those valid in the free theory and can be easily checked in perturbation theory at the first few orders of the 
weak coupling expansion. However, as we have stressed several times, they are valid for all values of the coupling constant and thus can be used even at strong coupling. It would be interesting to understand the field-theory origin of such identities which we have unveiled with our analysis.

Using these relations we have been able to write in full generality the structure constants in terms of the 2-point function, obtaining
\begin{equation}
    C_{U_k,T_\ell,\overbar{T}_p}=\frac{\sqrt{k}}{\sqrt{2}\,N}\,\sqrt{
    \ell+\lambda\partial_{\lambda}\big( \log G_{T_\ell} \big) }\,\sqrt{
    p+\lambda\partial_{\lambda}\big(\log G_{T_p} \big) }~,\label{CUTTwardconcl}
\end{equation}
and similarly for $C_{T_k,T_\ell,\overbar{U}_p}$.
From this relation and the strong-coupling expansion of the 2-point functions, it immediately follows that 
that the general structure of the structure constants at strong-coupling is that of an inverse power series of the shifted 't Hooft coupling
\begin{equation}
\sqrt{\lambda^\prime}=\sqrt{\lambda}-4\log 2~,
\label{lambdaprimebis}
\end{equation}
with coefficients proportional to odd Riemann $\zeta$-values and products thereof. In particular, we have found
\begin{equation}
C_{U_k,T_\ell,\overbar{T}_{p}} 
= C_{U_k,T_\ell,\overbar{T}_{p}}^{(\mathrm{LO})}
\,\Big(\frac{\lambda}{\lambda^\prime}\Big)^{\frac{1}{2}} 
\bigg[1+c_{\ell,p}^{(1)}\,\frac{\zeta_3}{\lambda^{\prime\,3/2}}+c_{\ell,p}^{(2)}\,\frac{\zeta_5}{\lambda^{\prime\,5/2}}+c_{\ell,p}^{(3)}\,\frac{\zeta_3^2}{\lambda^{\prime\,3}}+O\Big(\frac{1}{\lambda^{\prime\,7/2}}\Big)\bigg]+\cdots~,
\label{CUTTconcl}
\end{equation} 
and similarly for $C_{T_k,T_\ell,\overbar{U}_p}$.
In \cite{Billo:2022gmq,Billo:2022fnb} it was shown that the leading term $C_{U_k,T_\ell,\overbar{T}_{p}}^{(\mathrm{LO})}$ is reproduced exactly within the AdS/CFT correspondence by a holographic calculation using Type IIB supergravity in a $\mathbb{Z}_2$ orbifold background. Therefore, it is natural to expect that the sub-leading tail in the strong-coupling expansion (\ref{CUTTconcl}) arises from the string corrections to the supergravity effective action. The form of the terms in the square brackets of (\ref{CUTTconcl}) is consistent with this expectation. In fact the first string correction to the Type IIB supergravity effective action is proportional to $\alpha^{\prime\,3}\zeta_3$, where $\sqrt{\alpha^\prime}$ is the string length, while the subsequent ones are proportional to $\alpha^{\prime\,5}\zeta_5$, $\alpha^{\prime\,6}\zeta_3^2$ etc., as it can be checked by expanding the Type IIB closed string scattering amplitudes in powers of $\alpha^\prime$
(see for example \cite{Polchinski:1998rr}). On this basis, given the structure of (\ref{CUTTconcl}), it is highly tempting to speculate that the $\alpha^\prime$-expansion of the string amplitudes can be translated into an expansion in
inverse powers of $\sqrt{\lambda^\prime}$ or {\emph{vice-versa}} by means of the formal replacement
\begin{equation}
\alpha^\prime ~\longleftrightarrow~\frac{1}{\sqrt{\lambda^\prime}}
\label{speculation}
\end{equation}
in units where the AdS radius is 1.
To see whether this speculation has any basis, it should be shown that the coefficients 
$c_{\ell,p}^{(n)}$ in the square brackets of (\ref{CUTTconcl}) follow from the 2- and 3-point string amplitudes of the supergravity modes that are dual to the untwisted and twisted operators of the quiver theory. To compute these amplitudes a detailed knowledge of the higher-derivative interaction vertices induced by the string corrections in all sectors of the orbifold is required. This knowledge, however, is presently unavailable. Thus, as already pointed out in the introduction, the localization methods which we have described in this work are at the moment the only tools available to investigate the strong-coupling regime of the $\mathcal{N}=2$ quiver theory beyond leading order.

In our analysis we considered only the ``perturbative'' sub-leading terms proportional to inverse powers of the 't Hooft coupling but, as pointed out in \cite{Beccaria:2022ypy} for the 2-point functions, there are in addition non-perturbative corrections proportional to powers of $\exp(-\sqrt{\lambda})$ at large $\lambda$.
These exponentially small corrections result from the fact that the perturbative part of the strong-coupling expansion is non Borel summable due to the presence of singularities on the positive real axis of the Borel plane which require the addition of non-perturbative terms to fully determine the solution at strong coupling and its interpolation with the weak-coupling limit. It would be interesting to compute systematically also these non-perturbative corrections to the structure constants and investigate their holographic dual interpretation. It would also be nice to understand the meaning of the shift (\ref{lambdaprimebis}) and the origin of the factors of $(\lambda^\prime/\lambda)$ in the gravity side of the AdS/CFT correspondence.

Finally, we observe that our results can be easily generalized in several directions. For example, if we perform an orientifold projection of the quiver by identifying the fields of the two nodes as $\phi_0\leftrightarrow -\phi_1$, we obtain the $\mathbf{E}$ theory \cite{Billo:2019fbi,Beccaria:2021hvt}. With this action, the odd untwisted operators and the even twisted ones are projected out, but the surviving operators behave exactly as in the quiver theory, and thus their 2- and 3-point functions can be immediately read from the formulas we have presented in this work by only limiting the values of the conformal dimensions. Another generalization is to consider a quiver theory with $M>2$ nodes. In this case there are some simple modifications such as the replacement of the factors of $\sqrt{2}$ with $\sqrt{M}$ and the appearance of numerical coefficients in front of the $\mathsf{X}$ matrix that distinguish the different twisted sectors \cite{Billo:2022fnb}. The most important novelty, however, is the presence of non-trivial 3-point function with three twisted operators when the three twist parameters add to zero. In this case the structure constants have again the factorized structure of (\ref{CUTTwardconcl}) but with the factor $\sqrt{k}$ replaced by $\sqrt{k+\lambda\partial_{\lambda}\big(\log G_{T_k} \big) }$
for the appropriate third twisted operator. 

\vskip 1cm
\noindent {\large {\bf Acknowledgments}}
\vskip 0.2cm
We would like to thank Matteo Beccaria and Francesco Galvagno for useful discussions.
This research is partially supported by the MUR PRIN contract 2020KR4KN2 ``String Theory as a bridge between Gauge Theories and Quantum Gravity'' and by
the INFN project ST\&FI
``String Theory \& Fundamental Interactions''. 

\vskip 1cm

\appendix

\section{Proof of Eq.\texorpdfstring{\,(\ref{id2kkk0})}{}}
\label{appendix:A}

In this Appendix we prove the following identity:
\begin{align}
\label{id2kkk}
    \frac{1}{2}\sum_{n=2}^{k}\sum_{m=2}^{k} \mathsf{M}^-_{k,n}\,
    \mathsf{M}^-_{k,m} \big( n+m+2\lambda\partial_{\lambda} \big) T^{-,-}_{n,m}=
    \big( k+\lambda\partial_{\lambda} \big) \big\langle O_k^-\,O^-_k \big\rangle ~.
\end{align}
Let us first observe that using the definition of $O_k^-$ in (\ref{Mdef}) we have
\begin{align}
\label{2ptIdef}
\big\langle O^-_k\, O^-_k \big\rangle = 
\sum_{n=2}^k \sum_{m=2}^k \mathsf{M}^-_{k,n}\,\mathsf{M}^-_{k,m}	\,T^{-,-}_{n,m}~.
\end{align}
On the other hand, due to the orthogonality of operators with different dimensions, we have
\begin{align}
\label{2ptIIdef}
\big\langle O^-_k \,O^-_k \big\rangle 
&= \big\langle O^-_k \,A^-_k \big\rangle 
= \sum_{n=2}^k \mathsf{M}^-_{k,n}\,T^{-,-}_{n,k}~,\\
\label{ortho}
\big\langle O^-_k \,O^-_\ell \big\rangle &= \big\langle 
O^-_k \,A^-_\ell\big \rangle = \sum_{n=2}^k \mathsf{M}_{k,n}^-\,
T^{-,-}_{n,\ell}=0\qquad\mbox{for}~ \ell<k~.
\end{align}
We now consider the left-hand-side of \eqref{id2kkk} and concentrate on the first two terms which, taking into account that $\mathsf{M}_{k,k}^-=1$ and $T_{n,m}^{-,-}=T_{m,n}^{-,-}$, we can manipulate and rewrite as follows
\begin{align}
\frac{1}{2}\sum_{n=2}^{k}\sum_{m=2}^{k} \mathsf{M}^-_{k,n}\,
    \mathsf{M}^-_{k,m} \big( n+m\big) T^{-,-}_{n,m}  &=
    \sum_{n=2}^{k-1}\sum_{m=2}^{k}n\, \mathsf{M}^-_{k,n}\,
    \mathsf{M}^-_{k,m}\,T^{-,-}_{m,n}
    +k\sum_{m=2}^{k} \mathsf{M}^-_{k,m}\,T^{-,-}_{m,k}~.
\end{align}
In the first term of the right-hand-side, the sum over $m$ vanishes because of the orthogonality condition (\ref{ortho}) since $n<k$. Thus, we remain with the second term which is proportional to $\big\langle O^-_k \,O^-_k \big\rangle$ as one can see from (\ref{2ptIIdef}). In conclusion we have shown that
\begin{align}
\frac{1}{2}\sum_{n=2}^{k}\sum_{m=2}^{k} \mathsf{M}^-_{k,n}\,
    \mathsf{M}^-_{k,m} \big( n+m\big) T^{-,-}_{n,m}
= k\, \big\langle O_k^-\,O^-_k \big\rangle~.
\label{term1}
\end{align}
We finally consider the derivative term in the left-hand-side of \eqref{id2kkk}. Taking again into account the symmetry property $T_{n,m}^{-,-}=T_{m,n}^{-,-}$, we can write this term as
\begin{align}
 \sum_{n=2}^{k}\sum_{m=2}^{k} \mathsf{M}^-_{k,n}\,\mathsf{M}^-_{k,m} \,\lambda\partial_{\lambda} T^{-,-}_{n,m}=\sum_{n=2}^{k}\sum_{m=2}^{k} \lambda\partial_{\lambda} 
 \big(\mathsf{M}^-_{k,n}\,\mathsf{M}^-_{k,m} \,T^{-,-}_{n,m}\big) - 2
 \sum_{n=2}^{k}\sum_{m=2}^{k} \lambda\partial_{\lambda} \big(\mathsf{M}^-_{k,n}\big)
 \mathsf{M}^-_{k,m}\,T^{-,-}_{n,m}~.
\end{align}
The upper limit of sum over $n$ in the last term can be reduced to $k-1$, since when $n=k$ the mixing coefficient is $\mathsf{M}_{k,k}^-=1$ and its $\lambda$-derivative vanishes. Therefore, in this last term we have $n<k$ and the sum over $m$ vanishes because of the orthogonality condition (\ref{ortho}). Using 
\eqref{2ptIdef}, we conclude that
\begin{align}
 \sum_{n=2}^{k}\sum_{m=2}^{k} \mathsf{M}^-_{k,n}\,\mathsf{M}^-_{k,m}\, \lambda\partial_{\lambda} T^{-,-}_{n,m}=\lambda\partial_{\lambda} \big\langle O_k^-\,O^-_k \big\rangle~.
 \label{term2}
\end{align}
Adding (\ref{term1}) and (\ref{term2}) we obtain \eqref{id2kkk}, which is also Eq.\,(\ref{id2kkk0}) of the main text.

\section{Recursive procedure for \texorpdfstring{$\mathsf{D}_{k,\ell}$}{}}
\label{appendix:B}

In this appendix we discuss a recursive bootstrap-like procedure that allows us to obtain the strong-coupling expansion of the matrix elements $\mathsf{D}_{k,\ell}$. Even if this information is not strictly speaking necessary to derive the strong-coupling expansion of the 3-point functions and of the structure constants, we include this material for completeness. Moreover, it can be used also for an independent check on the results described in the main text.

To discuss the recursive procedure it is convenient to organize the matrix elements $\mathsf{D}_{k,\ell}$ according to their level, defined 
simply as $k+\ell$, as follows:
\begin{align}
&\bullet~\mbox{level 4}:\quad \mathsf{D}_{2,2}~,\notag\\
&\bullet~\mbox{level 6}:\quad \mathsf{D}_{3,3}~,~\mathsf{D}_{2,4}~,\label{table}\\
&\bullet~\mbox{level 8}:\quad \mathsf{D}_{4,4}~,~\mathsf{D}_{3,5}~,~\mathsf{D}_{2,6}~,\notag
\end{align}
and so on. There are two entries
for which we already know the strong-coupling expansion, namely 
$\mathsf{D}_{2,2}$ and $\mathsf{D}_{3,3}$. Indeed, as indicated in (\ref{Rkexp}), they are related to the ratios $R_2$ and $R_3$ respectively from which one finds
\begin{subequations}
\begin{align}
\mathsf{D}_{2,2}&=\frac{8\pi^2}{\lambda}
\Big(\frac{\lambda^\prime}{\lambda}\Big)\bigg[1+\frac{3\,\zeta_3}{\lambda^{\prime\,3/2}}+\frac{135\,\zeta_5}{16\lambda^{\prime\,5/2}}+\frac{81\,\zeta_3^2}{4\lambda^{\prime\,3}}+O\Big(\frac{1}{\lambda^{\prime\,7/2}}\Big)\bigg]~,\label{D22expanded}\\[1mm]
\mathsf{D}_{3,3}&=\frac{24\pi^2}{\lambda}
\Big(\frac{\lambda^\prime}{\lambda}\Big)^2\bigg[1+\frac{30\,\zeta_3}{\lambda^{\prime\,3/2}}-\frac{945\,\zeta_5}{8\lambda^{\prime\,5/2}}-\frac{405\,\zeta_3^2}{2\lambda^{\prime\,3}}+O\Big(\frac{1}{\lambda^{\prime\,7/2}}\Big)\bigg]~.
\label{D33expanded}
\end{align}
\label{D2233expanded}%
\end{subequations}
Thus, at level 4 there is nothing else to determine while at level 6 we have to find $\mathsf{D}_{2,4}$ and at all higher levels we have to determine everything. 
To this purpose we exploit a set of relations satisfied by $\mathsf{D}_{k,\ell}$ and their $\lambda$-derivatives which were presented in Appendix~A of \cite{Billo:2022fnb}. These relations, which involve the quantities $\mathsf{d}_k$ defined in (\ref{dk}), are\,%
\footnote{The identities \eqref{idEvenEven} and \eqref{idOddOdd} are the straightforward extensions for $n \geq 2$ of the identities (A.15) and (A.16) of \cite{Billo:2022fnb} respectively.}
\begin{align}
\mathsf{d}_{2n}\,\textrm{d}_{2m} &= \sqrt{2m}\sum_{r=1}^{m-1}\sqrt{2r} \,\mathsf{D}_{2r,2n} 
+ \sqrt{2n}\sum_{s=1}^{n-1}\sqrt{2s}\,\mathsf{D}_{2s,2m} + \big(n+m+\lambda \partial_{\lambda}\big)\mathsf{D}_{2n,2m}~, \label{idEvenEven}\\
\mathsf{d}_{2n+1}\,\mathsf{d}_{2m+1} &=\sqrt{2m+1}\sum_{r=1}^{m-1}\sqrt{2r+1}\, \mathsf{D}_{2r+1,2n+1} +\sqrt{2n+1}\sum_{s=1}^{n-1}\sqrt{2s+1}\, \mathsf{D}_{2s+1,2m+1}\nonumber \\
&\qquad\qquad + \big(n+m+1+\lambda \partial_{\lambda}\big)\mathsf{D}_{2n+1,2m+1}~, \label{idOddOdd}\\
\mathsf{d}_{2n}\,\mathsf{d}_{2m+1} &= \sqrt{2m+1}\sum_{r=1}^{m}\sqrt{2r} \, 
\mathsf{D}_{2r,2n} + \sqrt{2n}\sum_{s=1}^{n-1}\sqrt{2s+1} \, \mathsf{D}_{2s+1,2m+1}\ , \label{idEvenOdd}
\end{align}
for $n,m \geq 1$.
A few explicit examples, which will be useful in the following, are
\begin{subequations}
\begin{align}
\mathsf{d}_{2}\,\mathsf{d}_{2}&=2\,\mathsf{D}_{2,2}+\lambda\partial_\lambda\mathsf{D}_{2,2}~,
\qquad\quad\,\,\mathsf{d}_{3}\,\mathsf{d}_{3}=3\,\mathsf{D}_{3,3}+\lambda\partial_\lambda\mathsf{D}_{3,3}~,\label{id2233}\\[1mm]
\mathsf{d}_{2}\,\mathsf{d}_{3}&=\sqrt{6}\,\mathsf{D}_{2,2}~,\qquad\qquad\qquad\quad
\mathsf{d}_{2}\,\mathsf{d}_{4}=2\sqrt{2}\,\mathsf{D}_{2,2}+3\,\mathsf{D}_{2,4}+\lambda\partial_{\lambda}\mathsf{D}_{2,4}~,\label{id2324}\\[1mm]
\mathsf{d}_{3}\,\mathsf{d}_{4}&=\sqrt{6}\,\mathsf{D}_{2,4}+2\sqrt{3}\,\mathsf{D}_{3,3}~,\qquad
\mathsf{d}_{4}\,\mathsf{d}_{4}=4\sqrt{2}\,\mathsf{D}_{2,4}+4\,\mathsf{D}_{4,4}+\lambda\partial_{\lambda}\mathsf{D}_{4,4}~.\label{id3444}
\end{align}
\label{idexamples}%
\end{subequations}
To begin with, we use these relations to determine $\mathsf{D}_{2,4}$ that is the unknown element at level 6. It is straightforward to obtain
\begin{align}
\sqrt{6}\,\mathsf{D}_{2,2}&=\mathsf{d}_{2}\,\mathsf{d}_{3}
=\mathsf{d}_2\,\mathsf{d}_3\,\frac{\mathsf{d}_2\,\mathsf{d}_4}{\mathsf{d}_2\,\mathsf{d}_4}=
    \frac{(\mathsf{d}_2\,\mathsf{d}_2)\,(\mathsf{d}_3\,\mathsf{d}_4)}
    {\mathsf{d}_2\,\mathsf{d}_4} \notag\\
    &= \frac{\big(2\,\mathsf{D}_{2,2}+\lambda\partial_\lambda\mathsf{D}_{2,2}
    \big)\,\big(\sqrt{6}\,\mathsf{D}_{2,4}+2\sqrt{3}\,\mathsf{D}_{3,3}\big)}{2\sqrt{2}\,\mathsf{D}_{2,2}+3\,\mathsf{D}_{2,4}+\lambda\partial_{\lambda}\mathsf{D}_{2,4}} ~.
    \label{relationD24}
\end{align}
This is a nontrivial relation between $\mathsf{D}_{2,2}$ and $\mathsf{D}_{3,3}$, which are known, and $\mathsf{D}_{2,4}$, which is unknown. With elementary manipulations we can recast (\ref{relationD24}) in the
following form
\begin{align}
\mathsf{D}_{2,4}+\lambda\partial_\lambda \mathsf{D}_{2,4}-\mathsf{D}_{2,4}\,\frac{\lambda\partial_\lambda\mathsf{D}_{2,2}}{\mathsf{D}_{2,2}}=2\sqrt{2}\,\mathsf{D}_{3,3}-2\sqrt{2}\,\mathsf{D}_{2,2}+
\sqrt{2}\,\mathsf{D}_{3,3}\,\frac{\lambda\partial_\lambda\mathsf{D}_{2,2}}{\mathsf{D}_{2,2}}~.
\label{relationD24a}
\end{align}
If we write
\begin{align}
\label{D24}
    \mathsf{D}_{2,4} \ \underset{\lambda \rightarrow \infty}{\sim} \ 
    \sum_{n=1} \frac{d^{(n)}_{2,4}}{\lambda^{\frac{n+1}{2}}\phantom{\Big|}} \,=\, 
    \frac{d_{2,4}^{(1)}}{\lambda} + \frac{d_{2,4}^{(2)}}{\lambda^{3/2}} + \frac{d_{2,4}^{(3)}}{\lambda^{2}} + \cdots~,
\end{align}
and insert this expansion in (\ref{relationD24a}), we find a system of linear equations for the coefficients $c^{(n)}_{2,4}$. Solving the first equations of this system yields
\begin{align}
d_{2,4}^{(1)} = 8\sqrt{2}\pi^2~, \qquad d_{2,4}^{(2)} = -256\sqrt{2}\pi^2\log 2 ~.
\label{c2412}
\end{align}
The leading coefficient $d_{2,4}^{(1)}$ agrees of course with the result in (\ref{Dstrong}), while $d_{2,4}^{(2)}$ is a new result.
It is not difficult to realize that at order $1/\lambda^2$ the equation (\ref{relationD24a}) can be identically satisfied for any choice of $d_{2,4}^{(3)}$ which is therefore left undetermined. On the contrary, at higher orders one always finds consistency relations that fix all coefficients $d_{2,4}^{(n)}$ with $n\geq 4$ in terms of $d_{2,4}^{(3)}$.
In Table~\ref{tab_24} we provide the explicit expressions of these coefficients up to $n=7$.
\begin{table}[hbt!]
\centering
\begin{tabular}{|c||c|}
\hline 
\multicolumn{2}{|c|}{$\phantom{\Big|}~d_{2,4}^{(n)}$} \\ 
\hline
$\phantom{\big|}n=1$ & $8\,\sqrt{2}\,\pi^2$ \\  \hline
$\phantom{\big|}n=2$ & $-256\,\sqrt{2}\,\pi^2\log 2$ \\ \hline
$\phantom{\big|}n=3$ & $d_{2,4}^{(3)}$ \\ \hline
$\phantom{\big|}n=4$ & $10240\,\sqrt{2}\,\pi^2\log^3\!2-1056\,\sqrt{2}\,\pi^2\,\zeta_3-8\,\log 2\,
d_{2,4}^{(3)}$ \\ \hline
$\phantom{\big|}n=5$ & $-26624\,\sqrt{2}\,\pi^2 \log^4\!2
+3840\,\sqrt{2}\,\pi^2\log 2\,\zeta_3+16\,\log^2\!2\,d_{2,4}^{(3)}$ \\ \hline
$\phantom{\big|}n=6$ & $-6912\,\sqrt{2}\,\pi^2\log^2\!2\,\zeta_3
+2430\,\sqrt{2}\,\pi^2\,\zeta_5+3\,\zeta_3\,d_{2,4}^{(3)}$ \\ \hline
$\phantom{\big|}n=7$ & $-27648\,\sqrt{2}\,\pi^2\,\log^3\!2\,\zeta_3
+2592\,\sqrt{2}\,\pi^2\,\zeta_3^2+
8640\,\sqrt{2}\,\pi^2\,\log 2\,\zeta_5+12\,\log 2\,\zeta_3\,d_{2,4}^{(3)}$ \\ \hline
\end{tabular}
\caption{The first seven coefficients $d_{2,4}^{(n)}$ in the strong-coupling
expansion of $\mathsf{D}_{2,4}$.}
\label{tab_24}
\end{table}

Let us move to the next level and consider $\mathsf{D}_{4,4}$. Also in this case we can find a relation that allows us to determine the expansion of $\mathsf{D}_{4,4}$ in terms of those that we already know from the
previous analysis. Indeed, using the identities (\ref{idexamples}) in analogy
with (\ref{relationD24}), we have
\begin{align}
\sqrt{6}\,\mathsf{D}_{2,2}&=\mathsf{d}_{2}\,\mathsf{d}_{3}
=\mathsf{d}_2\,\mathsf{d}_3\,\frac{\mathsf{d}_4\,\mathsf{d}_4}{\mathsf{d}_4\,\mathsf{d}_4}=
    \frac{(\mathsf{d}_2\,\mathsf{d}_4)\,(\mathsf{d}_3\,\mathsf{d}_4)}
    {\mathsf{d}_4\,\mathsf{d}_4} \notag\\
    &= \frac{\big(2\sqrt{2}\,\mathsf{D}_{2,2}+3\,\mathsf{D}_{2,4}+\lambda\partial_{\lambda}\mathsf{D}_{2,4}\big)
  \,\big(\sqrt{6}\,\mathsf{D}_{2,4}+2\sqrt{3}\,\mathsf{D}_{3,3}\big) }{4\sqrt{2}\,\mathsf{D}_{2,4}+4\,\mathsf{D}_{4,4}+\lambda\partial_{\lambda}\mathsf{D}_{4,4}} ~.
    \label{relationD44}
\end{align}
In this relation everything is known except $\mathsf{D}_{4,4}$. To find its strong-coupling expansion we proceed as before and write
\begin{align}
\label{D44}
    \mathsf{D}_{4,4} \ \underset{\lambda \rightarrow \infty}{\sim} \ 
    \sum_{n=1} \frac{d^{(n)}_{4,4}}{\lambda^{\frac{n+1}{2}}\phantom{\Big|}} \,=\, 
    \frac{d_{4,4}^{(1)}}{\lambda} + \frac{d_{4,4}^{(2)}}{\lambda^{3/2}} + \frac{d_{4,4}^{(3)}}{\lambda^{2}} + \cdots~.
\end{align}
Inserting this expansion in (\ref{relationD44}), one can determine the coefficients $d_{4,4}^{(n)}$. Notice that since $\mathsf{D}_{4,4}$ appears in (\ref{relationD44}) only in the combination $\big(4\mathsf{D}_{4,4}+\lambda
\partial_\lambda\mathsf{D}_{4,4}\big)$, 
the coefficient $d_{4,4}^{(7)}$ of the term $1/\lambda^4$ cannot be fixed. One may think that in this way another ambiguity appears beside the one already observed for $d_{2,4}^{(3)}$ and that the number of free parameters grows in an uncontrollable way. However, this is not the case since there exists an independent relation that allows us to fix $d_{4,4}^{(7)}$. Indeed, from (\ref{Rkexp}) we see that $\mathsf{D}_{4,4}$
is related to $\mathsf{D}_{2,2}$ and $\mathsf{D}_{2,4}$ through the ration $R_4$, whose expansion is known. One can therefore check that the expansions (\ref{D22expanded}), (\ref{D24}) and (\ref{D44}) are consistent with that of $R_4$ if and only if 
\begin{align}
d_{4,4}^{(7)}&=64\,\pi^{2}\big(176128\,\log^6\!2-53760 \log^3\!2\,\zeta_3+1755\,\log 2\,\zeta_5
+3591\,\zeta_{3}^2 \big)\notag\\
&\quad
-64 \sqrt{2} \,\log 2\,\big(104\log^3\!2-15\,\zeta_3\big)\,c_{2,4}^{(3)}
+\frac{2\log^2\!2}{\pi^2}\,\big(d_{2,4}^{(3)}\big)^2~.
\end{align}
In this way all coefficients $d_{4,4}^{(n)}$, including $d_{4,4}^{(7)}$, are fully determined in terms of those appearing at the previous levels, and the only free parameter remains $d_{2,4}^{(3)}$. In Table~\ref{tab_44} we provide the explicit expression of the first few coefficients $d_{4,4}^{(n)}$ obtained in this way.
\begin{table}[hbt!]
\centering
\begin{tabular}{|c||c|}
\hline
\multicolumn{2}{|c|}{$\phantom{\Big|}~d_{4,4}^{(n)}$} \\
\hline
$\phantom{\big|}n=1$ & $64\,\pi^2$ \\ \hline
$\phantom{\big|}n=2$ & $-2048\,\pi^2\log 2$ \\ \hline
$\phantom{\big|}n=3$ &  $20480\,\pi^2\log^2\!2+2\,\sqrt{2}\,d_{2,4}^{(3)}$ \\ \hline
$\phantom{\big|}n=4$ & $65536\,\pi^2\log^3\!2+768\,\pi^2\,\zeta_3
-64\,\sqrt{2}\,\log 2\,d_{2,4}^{(3)}$ \\ \hline
$\phantom{\big|}n=5$ & $-360448\,\pi^2\log^4\!2+58368\,\pi^2\log 2\,\zeta_3+(d_{2,4}^{(3)})^2/(8\,\pi^2)$ \\ \hline
$n=6$ & $\makecell[l]{\qquad\qquad-3276800\,\pi^2\log^5\!2+724992\,\pi^2\log^2\!2\,\zeta_3-66960\,\pi^2\,\zeta_5\phantom{\Big|} ~~~~~\\
\qquad\qquad\quad+2560\,\sqrt{2}\,\log^3\!2\,d_{2,4}^{(3)} -264\,\sqrt{2}\,\zeta_3\,d_{2,4}^{(3)}-\log 2\,(d_{2,4}^{(3)})^2/\pi^2}$ \\
\hline
$n=7$ & $\makecell[l]{
~11272192\,\pi^2 \log^6\!2-3440640\,\pi^2\log^3\!2\,\zeta_3+112320\,\pi^2\log 2\,\zeta_5+229824\,\pi^2\,\zeta_3^2\phantom{\Big|}~
\\
\qquad\quad
-6656 \,\sqrt{2}\log^4\!2\,d_{2,4}^{(3)}+960\,\sqrt{2}\log 2\,\zeta_3\,d_{2,4}^{(3)}
+2\log^2\! 2\,(d_{2,4}^{(3)})^2/\pi^2}$ 
\\ \hline
\end{tabular}
\caption{The first seven coefficients $d_{4,4}^{(n)}$ in the strong-coupling
expansion of $\mathsf{D}_{4,4}$.}
\label{tab_44}
\end{table}

Let us now consider the next entry at level 8, namely $\mathsf{D}_{3,5}$. Its
strong-coupling expansion
\begin{align}
\label{D35}
    \mathsf{D}_{3,5} \ \underset{\lambda \rightarrow \infty}{\sim} \ 
    \sum_{n=1} \frac{d^{(n)}_{3,5}}{\lambda^{\frac{n+1}{2}}\phantom{\Big|}} 
\end{align}
can be fully determined using the following relation
\begin{align}
\sqrt{6}\,\mathsf{D}_{2,2}&=\mathsf{d}_{2}\,\mathsf{d}_{3}
=\mathsf{d}_2\,\mathsf{d}_3\,\frac{\mathsf{d}_4\,\mathsf{d}_5}{\mathsf{d}_4\,\mathsf{d}_5}=
    \frac{(\mathsf{d}_2\,\mathsf{d}_5)\,(\mathsf{d}_3\,\mathsf{d}_4)}
    {\mathsf{d}_4\,\mathsf{d}_5} \notag\\
    &= \frac{\big(\sqrt{10}\,\mathsf{D}_{2,2}+2\sqrt{5}\,\mathsf{D}_{2,4}\big)
  \,\big(\sqrt{6}\,\mathsf{D}_{2,4}+2\sqrt{3}\,\mathsf{D}_{3,3}\big) }{2\sqrt{3}\,\mathsf{D}_{3,5}+\sqrt{10}\,\mathsf{D}_{2,4}+2\sqrt{5}\,\mathsf{D}_{4,4}} ~.
    \label{relationD35}
\end{align}
Since here no $\lambda$-derivatives of $\mathsf{D}_{3,5}$
appear, all coefficients $d^{(n)}_{3,5}$ can be fixed without introducing any new free parameter. The first few of such coefficients are listed in Table~\ref{tab_35}.
\begin{table}[hbt!]
\centering
\begin{tabular}{|c||c|}
\hline
\multicolumn{2}{|c|}{$\phantom{\Big|}~d_{3,5}^{(n)}$} \\
\hline
$\phantom{\big|}n=1$ & $8\,\sqrt{15}\,\pi^2$ \\ \hline
$\phantom{\big|}n=2$ & $-384\,\sqrt{15}\,\pi^2\log 2$ \\ \hline
$\phantom{\big|}n=3$ & $1280\,\sqrt{15}\,\pi^2\log^2\!2+\sqrt{30}\,d_{2,4}^{(3)}$ \\ \hline
$\phantom{\big|}n=4$ & $30720\,\sqrt{15}\,\pi^2\log^3\!2-3120\,\sqrt{15}\,\pi^2\,\zeta_3-16\,\sqrt{30}\,\log 2\,d_{2,4}^{(3)}$ \\ \hline
$\phantom{\big|}n=5$ & $-276480\,\sqrt{15}\,\pi^2\log^4\!2+31680\,\sqrt{15}\,\pi^2\log 2\,\zeta_3+96\,\sqrt{30}\,\log^2\!2\,d_{2,4}^{(3)}$ \\ \hline
$n=6$ & $\makecell[l]{\quad\qquad 851968\,\sqrt{15}\,\pi^2\log^5\!2-238080\,\sqrt{15}\,\pi^2\log^2\!2\,\zeta_3+27405\,\sqrt{15}\,\pi^2\,\zeta_5 \phantom{\Big|}\\ 
\qquad\qquad\qquad~~-256\,\sqrt{30}\,\log^3\!2\,d_{2,4}^{(3)}+30\,\sqrt{30}\,\zeta_3\,d_{2,4}^{(3)}}$ \\
\hline
$n=7$ & $\makecell[l]{~~-917504 \,\sqrt{15}\,\pi^2\log^6\!2+645120 \,\sqrt{15}\,\pi^2\log^3\!2\,\zeta_3-86940\,\sqrt{15}\,\pi^2\log 2\,\zeta_5 \phantom{\Big|}~~~~\\ 
\qquad~~~-34380\,\sqrt{15}\,\pi^2\,\zeta_3^2+256\,\sqrt{30}\,\log^4\!2\,d_{2,4}^{(3)}-120\,\sqrt{30}\,\log 2\,\zeta_3\,d_{2,4}^{(3)}}$ \\
\hline
\end{tabular}
\caption{The first seven coefficients $d_{3,5}^{(n)}$ in the strong-coupling expansion of $\mathsf{D}_{3,5}$.}
\label{tab_35}
\end{table}

The last entry at level 8 is $\mathsf{D}_{2,6}$. As before, also in this case we can find a nontrivial relation among the $\mathsf{D}_{k,\ell}$ already determined which allows us to fix the coefficients of the
strong-coupling expansion of $\mathsf{D}_{2,6}$. However, the coefficient
$d_{2,6}^{(5)}$ is left unconstrained by this relation and thus a new free parameter beside $d_{2,4}^{(3)}$ appears at this stage. 

\begin{figure}[ht]
    \centering
    \includegraphics[scale=0.65]{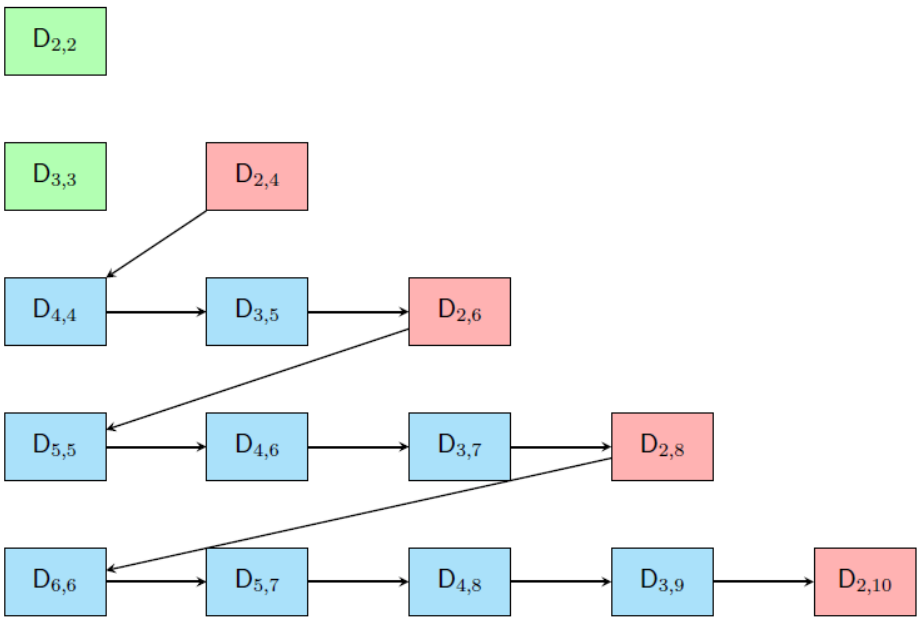}
    \caption{The sequence of steps that allow us to deduce the strong coupling expansions of $\mathsf{D}_{k,\ell}$ starting from those of $\mathsf{D}_{2,2}$ and $\mathsf{D}_{3,3}$ that are obtained using the results of \cite{Beccaria:2022ypy} and are given in (\ref{D2233expanded}). In blue we have indicated the entries that can be completely determined using the information obtained in the preceding steps of the procedure, while in red we have indicated the elements that at each level introduce a new free parameter.}
    \label{fig:levels}
\end{figure}

This constructive procedure can be iterated at the next level 
where, using relations of the type (\ref{relationD24}) and (\ref{relationD44}) combined with the 2-point functions (\ref{GTquater}), one finds that the strong-coupling expansions of $\mathsf{D}_{5,5}$, $\mathsf{D}_{4,6}$ and $\mathsf{D}_{3,7}$ are fully determined by the expansions of $\mathsf{D}_{k,\ell}$ at the previous levels, while the one of $\mathsf{D}_{2,8}$ depends on a new free
parameter, namely the coefficient $d_{2,8}^{(7)}$ which remains unconstrained.
We have explicitly verified that this structure replicates at the next levels and that, in general, the coefficients of the type $d_{2,2n}^{(2n-1)}$ are free parameters in the strong-coupling expansions. The recursive procedure that we have described above is schematically represented in Fig.~\ref{fig:levels} where the arrows indicate the sequence of steps that one has to take in order to progressively reconstruct $\mathsf{D}_{k,\ell}$ starting from the initial information contained in $\mathsf{D}_{2,2}$ and $\mathsf{D}_{3,3}$.

A few observations are in order. Firstly, we observe that differently from the expansions of the 2- and 3-point functions and of the structure constants presented in the main text, in the expansions of $\mathsf{D}_{k,\ell}$ one can not resum the $\log 2$ factors and exhibit the result as a series in inverse powers of the shifted coupling $\lambda^\prime$ defined in (\ref{lambdaprime}). Secondly, the coefficient $d_{3,5}^{(3)}$ has been obtained in \cite{Bobev:2022grf}
with a numerical approach (see Eq.~(3.25) of \cite{Bobev:2022grf} for $k=1$ and $\ell=2$). If we match this result with our findings (see the $n=3$ entry of Table~\ref{tab_35}) we can fix the free parameter $d_{2,4}^{(3)}$ and get
\begin{align}
 d^{(3)}_{2,4}= 64\sqrt{2}\,\pi^2\big(\pi^2+32\log^2\! 2\big)~.
 \label{c24fixed}
\end{align}
Using this expression, all coefficients of $1/\lambda^2$ in the expansion of $\mathsf{D}_{k,\ell}$ are fully determined and the ones of $\mathsf{D}_{2n+1,2m+1}$ match those of \cite{Bobev:2022grf}.

The above results can be used to obtain the strong-coupling expansion of the 3-point functions in a different way than discussed in Section~\ref{sec:general}. In fact, as explained in \cite{Billo:2022fnb}, it is always possible to write a 3-point function as a combination $\mathsf{D}_{k,\ell}$ and their $\lambda$-derivatives.
For example one has
\begin{align}
\label{234final}
G_{U_2,T_2,\overbar{T}_4} = \mathcal{G}_{2,2,4}\bigg[\mathsf{D}_{2,2}+\frac{1}{\sqrt{8}}\,\mathsf{D}_{2,4} + \frac{1}{\sqrt{8}}\,\lambda\, \partial_{\lambda}\Big(\frac{\mathsf{D}_{2,4}}{\mathsf{D}_{2,2}}\Big)\,\mathsf{D}_{2,2}\bigg]
\end{align}
Therefore, using (\ref{D22expanded}) and the expansion of $\mathsf{D}_{2,4}$ obtained with the recursive procedure, one finds
\begin{align}
G_{U_2,T_2,\overbar{T}_4}=\mathcal{G}_{2,2,4}\,
\frac{12 \pi ^2}{\lambda}\Big(\frac{\lambda^\prime}{\lambda}\Big)^{\frac{3}{2}} 
\bigg[1+\frac{51\,\zeta_3}{2\lambda^{\prime\,3/2}}-\frac{4455\,\zeta_5}{32\lambda^{\prime\,5/2}}-\frac{1539\,\zeta_3^2}{4\lambda^{\prime\,3}}+O\Big(\frac{1}{\lambda^{\prime\,7/2}}\Big)\bigg]~,
\label{GUTT224}
\end{align}
which agrees with (\ref{GUTTexpanded}) for $k=\ell=2$ and $p=4$.
Notice that all dependence on the free parameter $d_{2,4}^{(3)}$ that is present 
in the expansion of
$\mathsf{D}_{2,4}$ drops out in the 3-point function. This is so because in (\ref{234final}) the terms which involve $\mathsf{D}_{2,4}$ and could bring a dependence on $d_{2,4}^{(3)}$, are proportional to 
\begin{align}
\mathsf{D}_{2,4}+\lambda\partial_\lambda \mathsf{D}_{2,4}-\mathsf{D}_{2,4}\,\frac{\lambda\partial_\lambda\mathsf{D}_{2,2}}{\mathsf{D}_{2,2}}~.
\label{D24terms}
\end{align}
However, this is precisely the combination appearing in the left-hand-side of the identity (\ref{relationD24a}) which expresses it in terms of $\mathsf{D}_{2,2}$ and $\mathsf{D}_{3,3}$.
This explains why the free parameter $d_{2,4}^{(3)}$ does not appear in the final result (\ref{GUTT224}).

We have computed in this way several 3-point functions. Of course the calculations become increasingly more involved when the conformal dimensions of the operator become large, but in all case we have examined we have always found that the undetermined parameters $d_{2,2n}^{(2n-1)}$ that appear in the expansions of $\mathsf{D}_{k,\ell}$ always cancel in the final expressions and that the results are in full agreement with those found with the more economic approach discussed in the main text. This agreement can be seen also a
check on the validity of the recursive procedure we have illustrated in this appendix.

%
%
%

\providecommand{\href}[2]{#2}\begingroup\raggedright\endgroup

\end{document}